\documentstyle[aps,pre,epsf,eqsecnum]{revtex}
\begin{document}
%\narrowtext
\draft
%\tighten
\title
{
Soliton approach to the noisy Burgers equation \\ 
Steepest descent method
}
\author{Hans C. Fogedby}
\address{
\thanks{Permanent address}
Institute of Physics and Astronomy,
University of Aarhus, DK-8000, Aarhus C, Denmark\\
and\\
NORDITA, Blegdamsvej 17, DK-2100, Copenhagen {\O}, Denmark
}
\date{\today}
\maketitle
\begin{abstract}
The noisy Burgers equation in one spatial dimension is analyzed by
means of the Martin-Siggia-Rose technique in functional form.
In a canonical formulation the morphology and scaling behavior
are accessed by mean of a principle of least action in the
asymptotic non-perturbative weak noise limit. The ensuing coupled
saddle point field equations for the local slope and noise fields,
replacing the noisy Burgers equation, are solved yielding nonlinear
localized soliton solutions and extended linear diffusive mode solutions,
describing the morphology of a growing interface. The canonical formalism
and the principle of least action also associate momentum, energy, and action
with a soliton-diffusive mode configuration and thus provides a
selection criterion for the noise-induced fluctuations. In a
``quantum mechanical'' representation of the path integral the noise
fluctuations, corresponding to different paths in the path integral,
are interpreted as ``quantum fluctuations'' and the growth morphology
represented by a Landau-type quasi-particle gas of ``quantum solitons''
with gapless dispersion $E\propto P^{3/2}$ and ``quantum diffusive
modes'' with a gap in the spectrum. Finally, the scaling properties
are dicussed from a heuristic point of view in terms of a``quantum
spectral representation'' for the slope correlations. The dynamic
eponent $z=3/2$ is given by the gapless soliton dispersion law,
whereas the roughness exponent $\zeta =1/2$ follows from a
regularity property of the form factor in the spectral representation.
A heuristic expression for the scaling function is given by spectral
representation and has a form similar to the probability distribution
for L\'{e}vy flights with index $z$.
\end{abstract}
%\vfill
%E-mail:\\
%$1$. fogedby@dfi.aau.dk\\
%\newpage
%\narrowtext
%%%%%%%%%%%
%%%%%%%%%%%
%%%%%%%%%%%
%%%%%%%%%%%
%%%%%%%%%%%
%%%%%%%%%%%
\section{Introduction}
%%%%%%%%%%%
%%%%%%%%%%%
%%%%%%%%%%%
%%%%%%%%%%%
%%%%%%%%%%%
%%%%%%%%%%%
This is the second of a series of papers where we analyze the Burgers
equation in one spatial dimension with the purpose of modelling
the growth of an interface; for a brief account we refer to
\cite{fogedby1}. In the first paper, denoted in the following
by A \cite{fogedby2}, we investigated the {\em noiseless} Burgers equation 
\cite{burgers,saffman,jackson,whitham}
in terms of its nonlinear
soliton or shock wave excitations and
linear diffusive modes.
In the present paper we address our main objective, namely the {\em noisy} 
Burgers equation in one
spatial dimension \cite{forster}. This equation has the form
\begin{equation}
\frac{\partial u}{\partial t} = \nu\nabla^2 u + \lambda u \nabla u
+\nabla\eta  ~ ,
\label{bur}
\end{equation}
where $\nu$ is a damping constant or viscosity and $\lambda$ a nonlinear
coupling strength. The equation is driven by a conserved white noise
term, $\nabla\eta$, where $\eta$ has a Gaussian distribution and 
is short-range correlated in space
according to 
\begin{equation}
\langle\eta(x,t)\eta(x',t')\rangle
=
\Delta\delta(x-x')\delta(t-t') ~ .
\label{noise}
\end{equation}
%.
In the context of modelling a growing interface
the Kardar-Parisi-Zhang equation (KPZ)
for the height field $h$ \cite{kardar1}
\begin{equation}
\frac{\partial h}{\partial t} = \nu\nabla^2 h + 
\frac{\lambda}{2}(\nabla h)^2
+\eta ~ ,
\label{kpz}
\end{equation}
is equivalent to the Burgers equation
by means of the relationships 
\begin{eqnarray}
u =&& \nabla h 
\label{slope}
\\
h =&& \int udx ~ ,
\label{height}
\end{eqnarray}
that is the Burgers equation governs the dynamics of the  local slope 
of the interface. In Fig. 1 we have sketched the growth morphology in terms
of the height and slope fields.

The substantial conceptual problems encountered in nonequilibrium physics
are in many ways embodied in the Burgers-KPZ equations (\ref{bur}) and
(\ref{kpz})
which describe the self-affine growth of an interface subject to 
annealed noise arising from fluctuations in the drive or in 
the environment
\cite{halpin,barabasi,krug1,family1,beijeren1,kardar2,krug2}.
Interestingly, the Burgers-KPZ equations are also encountered in a variety
of other problems such as randomly stirred fluids
\cite{forster}, dissipative transport in a driven lattice gas
\cite{beijeren2,janssen1,ligget}, the propagation of flame fronts 
\cite{lvov,procaccia,zalesky}, 
the sine-Gordon equation \cite{krug3}, and magnetic
flux lines in superconductors\cite{hwa}. Furthermore, by means of the Cole-
Hopf transformation\cite{cole,hopf} the Burgers-KPZ equations are also 
related to
the problem of a directed polymer \cite{huse1,huse2} or 
a quantum particle in a random
medium \cite{kardar3,kardar4} and thus to the theory of spin glasses 
\cite{fisher,parisi,mezard}.

In contrast to the case of the noiseless Burgers equation discussed in A 
where the slope field
eventually relaxes due to the dissipative term $\nu\nabla^2u$, unless
energy is supplied to the system at the boundaries, the noisy Burgers equation
(\ref{bur}) describes an open nonlinear dissipative system driven 
into a stationary state with random energy input at a short wave length
scale provided by the conserved noise, $\nabla\eta$.
In the stationary regime the equation
thus describes time-independent stochastic self-affine roughened growth.
In the linear case for $\lambda = 0$ the Burgers equation reduces to the 
noise-driven  Edwards-Wilkinson equation (EW)
\cite{ew}
\begin{equation}
\frac{\partial u}{\partial t} = \nu\nabla^2u+\nabla\eta  ~ ;
\label{ew1}
\end{equation}
here for the slope field $u$. 
Owing to the absence of the nonlinear 
growth term $\lambda u\nabla u$ the cascade in wave number space
is absent and the correlations, probability distributions, 
and scaling properties are easy to derive
\cite{nattermann}. Furthermore, unlike the Burgers case 
the  EW equation, 
being compatible with
a fluctuation-dissipation theorem, actually describes the dynamic fluctuations
in an equilibrium state with temperature $\Delta/2\nu$ (in units such
that $k_B=1$) and as a consequence does not provide a proper description 
of a growing interface. 
On the other hand, the presence of the nonlinear growth term 
$\lambda u\nabla u$ 
in Eq. (\ref{bur}) renders it much more complicated and much richer.
The term 
filters the
input noise $\nabla\eta$ and gives rise to interactions between different
wave number components leading to a cascade which changes both the
scaling
properties and the probability distributions from the linear EW case.

The Burgers-KPZ equations owing to their simple form
accompanied by their very complex behavior have served as  paradigms 
in the theory
of driven and disordered systems and have been studied intensively
\cite{halpin,barabasi,krug1,family1,beijeren1,kardar2,krug2}.
One set of issues which have been addressed
are the scaling properties \cite{ma,chaikin,plischke,family2}.
According to the dynamic scaling hypothesis \cite{ma,family2}
supported by numerical simulations
and the fixed point structure of a renormalization group scaling analysis,
\cite{forster,kardar1},
the slope field $u$ is statistically scale invariant in the sense that the
self-affine rescaled $u'(x,t) = b^{-(\zeta -1)}u(bx,b^zt)$ is statistically
equivalent
to $u(x,t)$, where $b$ is a scale parameter. More precisely, the scaling
hypothesis implies the following dynamical scaling form for the 
slope correlation function in the stationary regime 
\cite{forster,kardar1,family2,family3,jullien}: 
\begin{equation}
\langle u(x,t)u(x,'t')\rangle = |x-x'|^{2(\zeta -1)}f(|t-t'|/|x-x'|^z) ~ .
\label{sf}
\end{equation}
The scaling behavior in the long wave length-low frequency limit is 
thus governed by two scaling dimensions: i) the roughness or wandering exponent
$\zeta$,  characterizing the slope correlations for a stationary profile
and ii) the dynamic exponent $z$, describing the temporal scaling in the
stationary regime \cite{footnote1}. 
The slope
field $u$ has the scaling dimension $1-\zeta$. For large $w$ the scaling
function $f(w)\propto w^{-2(1-\zeta)/z}$; for small $w$ 
$f(x)\propto \text{const}$.

Two properties determine the scaling exponents, namely a scaling law
and an effective fluctuation-dissipation theorem.
Like the noiseless or deterministic Burgers equation 
discussed in A, 
the noisy equation is also invariant under
a nonlinear 
Galilean transformation
\cite{forster,kardar1} 
\begin{eqnarray}
x&&\rightarrow x-\lambda u_0t
\label{galx}
\\
u&&\rightarrow u+u_0 ~ .
\label{galu}
\end{eqnarray}
Since the nonlinear coupling strength $\lambda$ here enters as a structural 
constant of the Galilean symmetry group  it transforms 
trivially under a
scaling transformation  and we infer the scaling law 
\cite{forster,kardar1,halpin,barabasi} 
\begin{equation}
\zeta + z = 2 ~ ,
\label{sl}
\end{equation}
relating $\zeta$ and $z$.
Furthermore, noting that the stationary Fokker-Planck equation for the Burgers
equation (\ref{bur})
is solved by a  Gaussian distribution \cite{forster,krug1,huse2}
\begin{equation}
P(u)\propto\exp{\left[-\frac{\nu}{\Delta}\int dx u^2\right]} ~ ,
\label{gauss}
\end{equation}
{\em independent} of $\lambda$ it follows that $u$ is an independent
random variable and that the height variable $h$ according to Eq. (\ref{height})
performs random walk, corresponding to the roughness exponent $\zeta=1/2$,
also in the linear EW case.
From the scaling law (\ref{sl}) we subsequently  obtain
the dynamic exponent $z=3/2$. In the linear EW case $\zeta=1/2$ and $z=2$,
characteristic of diffusion. In Table 1 we have summarized the exponents
for the EW and Burgers/KPZ universality classes.
\begin{center}
\begin{tabular}{||c||c|c||}
\hline
Universality class~~~~	&$~~~~~~\zeta~~~~~~$	&$~~~~~~~z~~~~~~~$\\
\hline
EW			&$1/2$	&2	 	\\
\hline
Burgers-KPZ		&$1/2$  &$3/2$	\\
\hline
\end{tabular}
\end{center}
\centerline{Table 1. Exponents and universality classes}
\smallskip
\smallskip
\smallskip

The standard tool used in the analysis of the scaling properties of nonlinear
Langevin equations of the type in Eqs. (\ref{bur}-\ref{kpz}) 
is the dynamic
renormalization group (DRG) method  \cite{ma}. The method is
based on an expansion in powers of the nonlinear couplings, an
expansion in the noise strength, arising from the noise contractions 
when implementing the statistical average, combined with an infinitesimal
momentum shell integration in the short wave length limit, corresponding 
to the scaling transformation.
The method operates in
variable dimension $d$ and typically identifies critical dimensions
separating regions where infrared convergent perturbation theory holds
yielding mean field behavior from regions with infrared divergent 
expansions. Here the DRG
allows for an organisation of the divergent terms and yields 
renormalization group equations for the effective parameters in
the theory in terms of an epsilon expansion
about the critical dimension . In powers of epsilon the 
DRG thus yields expression for the critical exponents
and information about the scaling functions.

For the Burgers-KPZ equations the expansion is in powers 
of the nonlinear coupling strength $\lambda$ and the noise strength
$\Delta$. The critical dimension is $d_c = 2$, also following from simple power
counting. Below $d_c=2$ there appear two renormalization group fixed points:
An unstable Gaussian fixed point, corresponding to vanishing coupling 
strength, describing a smooth interface governed by
the EW equation, and a stable strong coupling fixed point, characterizing 
a rough interface. The exponents assume non-trivial values for all
$\lambda\neq 0$. In  $d=1$ an effective fluctuation-dissipation
theorem equivalent to the Gaussian form in Eq. (\ref{gauss}) is operative,
and  together with the Galilean invariance,
implying trivial scaling of $\lambda$, the renormalization group equations 
yield the exponents in Table 1 associated with
the infrared stable non-trivial strong coupling fixed  point
\cite{forster,kardar1,mazenko,frey1,tauber}.
Unlike the case of static and dynamic critical phenomena, where the
renormalization group methods have proven very successful
\cite{ma,binney}, the situation
in the case of nonequilibrium phenomena  exemplified here by
the Burgers-KPZ equations has proven more difficult and despite
extensive efforts based on DRG calculations to higher
loop order \cite{frey1,tauber} and mode coupling approaches
\cite{frey2} the physics of the strong coupling fixed point in $d=1$ still 
remains elusive.

In a recent letter, denoted in the following by L \cite{fogedby3}, 
we approached the strong coupling fixed point behavior 
from the point of view of the mapping 
of the Burgers equation onto 
an equivalent solid-on-solid or  driven lattice gas  model
\cite{krug1,stinchcombe}, which furthermore maps onto a discrete spin 
$1/2$ chain model
\cite{mattis,fogedby4}.
The quantum spin chain approach has been proposed in
\cite{krug1,gwa}, and
considered further in \cite{neergard}, on the basis of the equivalence
between the Liouville operator in the Master
equation describing the evolution of the  one-dimensional
driven lattice gas, or the
equivalent lattice interface solid-on-solid growth model, and a non-Hermitian
spin $1/2$  Hamiltonian.
The quantum chain model has been treated by 
means of Bethe-Ansatz methods \cite{krug1,gwa,dahr,neergard,kim}
and the dynamic exponent $z=3/2$ obtained from
the finite size mass gap scaling. 
In L we pushed the analysis further and constructing a harmonic oscillator 
representation valid for large spin in combination with
a continuum limit  we derived a Hamiltonian
description and a set of coupled field equations of motion for the spin field,
corresponding to the slope  $u$, and a conjugate ``azimuthal'' 
angle field  replacing the noise. The field equations admit spin
wave solutions, corresponding to the linear diffusive modes and,
more importantly, nonlinear localized soliton solutions,
describing the growing steps in the original KPZ equation
or the solitons or shocks in the Burgers equation. We also derived the 
soliton dispersion
law and after a quasi-classical quantization identified in a heuristic
manner the elementary excitations of the theory. From the dispersion law
we deduced the dynamic
exponent $z=3/2$, characteristic of the zero temperature fixed point
of the ``quantum theory''. 
The picture that emerged from our analysis was 
that of a dilute quasi-particle gas of nonlinear soliton modes yielding
$z=3/2$ and a superposed  spin wave gas, corresponding to $z=2$, the
dynamic exponent for the linear case. In L we also
briefly discussed the operator algebra associated with the Hamiltonian 
representation and
derived the field equations by means of a canonical representation of
the Fokker-Planck equation for the equivalent Burgers equation. Whereas
the Bethe-Ansatz investigations by their nature are restricted to special
values of the coupling strength, corresponding to the fully asymmetric
exclusion model \cite{stinchcombe}, our analysis is valid  for general coupling strength
and thus constitutes an extension of the Bethe-Ansatz method to the
general case of a continuum field theory. The analysis in L was in
many respects incomplete and preliminary but it did indicate that
the strong coupling fixed point behavior is intrinsically
associated with the soliton modes in the Burgers equation
since they
both provide aspects of the growth morphology and also, independently,
yield the dynamic exponent.

Here we present a {\em unified} approach to the noisy Burgers
based on the Martin-Siggia-Rose technique in functional form
\cite{msr,dedom1,dedom2,dedom3,janssen2,baussch}. 
This method supersedes the analysis in L and does not make use of the 
mapping to a 
spin chain via a solid-on-solid model  and therefore the implicit 
assumption of 
persistent universality classes
under the mappings. Clearly, the different formulations are basically
equivalent arising as they do from the same basic stochastic growth
problem. The equivalence also indirectly demonstrates that the
scaling properties of the different models, i.e., the solid-on-solid model,
the spin chain, and the field formulation, fall in the same
KPZ-Burgers universality class. The present functional path integral method
which also can be seen as a generalization of the stationary distribution
(\ref{gauss}) to the time-dependent nonlinear case
provides a many-body description of the morphology of
a growing interface in terms of soliton excitations and does also give 
insight into the scaling behavior. Below we highlight some of our results.
\begin{itemize}
\item
The path integral formulation yields
a compact description of the noisy Burgers equation and provides
expressions for the probability distributions and correlation
functions. Reformulated as a canonical Feynman-type phase space 
path integral the approach allows for 
{\em a principle of least action}. Hence the weight of the different paths 
or interface
configurations, corresponding to the noise-induced interface fluctuations,
contributing to the path integral are 
controlled by an effective action. The role of the effective
Planck constant is here played by the noise correlation strength $\Delta$.
The action in the path integral thus plays the same role 
for the dynamical configurations as the Hamiltonian
in the Boltzmann factor for the static configurations in equilibrium 
statistical mechanics.
\item
In the asymptotic weak noise limit the principle of least action
implies that the dominant configurations arising from the
solutions of the  saddle point field equations correspond to a
a dilute nonlinear soliton gas with superposed linear diffusive modes.
The canonical formulation and the principle of least action 
furthermore allow a dynamical description and associate energy, 
momentum, and action with
the solitons and the diffusive modes.
\item
The path integral formulation permits a ``quantum mechanical''
interpretation in terms of an underlying non-Hermitian relaxational
``quantum mechanics'' or ``quantum field theory''. The noise-induced  
fluctuations here
correspond to 
``quantum fluctuations'' and the fluctuating growth morphology 
is described by a Landau-type quasi-particle gas of nonlinear 
``quantum solitons''
and linear ``quantum diffusive modes''.
In the height field this corresponds to a morphology
of growing steps  with superposed linear modes.
The ``quantum soliton'' dispersion law is gapless and characterized
by an exponent $3/2$; the ``quantum diffusive mode'' dispersion law
is quadratic with a gap in the spectrum proportional to the soliton
amplitude.
\item
In the present formulation the scaling properties associated
with the ``zero temperature'' fixed point in the underlying
``quantum field theory'' follow as a by-product from the soliton
and diffusive mode dispersion laws and the spectral representation
of the correlations. 
The dominant excitation in the long wave length-low frequency limit
identifies the relevant universality class.
The present many-body formulation yields
the known exponents. The dynamic exponents $z=3/2$ and $z=2$ are 
associated with
the soliton and diffusive mode dispersion laws, respectively,
whereas the  roughness exponent  $\zeta =1/2$ follows from a regularity
property of the form factor in the spectral representation.
The many-body formulation also explains the robustness
of the roughness exponent under a change of universality class
and provides 
a heuristic expression for the scaling function
which has the same structure as the probability distribution 
for L\'{e}vy flights.
\item
From a field theoretical point of view we identify the noise strength
$\Delta$  as the effective {\em small parameter}. Furthermore,
the fundamental probability distribution or path integral
has an {\em essential singularity}
for $\Delta = 0$.
Hence our approach is based on a {\em non-perturbative
saddle point or steepest descent approximation} to the path integral.
We believe that it is  in this respect that the dynamic renormalization group
method based on an expansion in $\lambda$ {\em and} in the noise contraction
$\Delta$ fails to access the strong coupling fixed point.
\end{itemize}

The path integral representation of the noisy 
Burgers equation presented here is equivalent to a full-fledged 
one dimensional non-Hermitian
non-Lagrangian field theory and requires for its detailed analysis
some advanced  field theoretical techniques and methods
from quantum chaos.
In the present context we choose, however, a somewhat
heuristic approach to the path integral in order to elucidate the emerging
simple physical picture of a growing interface. This approach then also 
serves as
a tutorial introduction to the field theoretical treatment 
to be presented elsewhere.

The present paper is organized in the following way. In section II we
discuss the simple case of the linear Edwards-Wilkinson
equation, mainly in order to emphasize the non-perturbative nature
of the noise as regards the stationary driven regime. Since the soliton
modes in the noisy Burgers equation turn out to be of crucial importance 
in understanding the morphology and scaling properties, we summarize
in section III the results obtained in A concerning the solitons
and diffusive modes in the noiseless Burgers equation. In section IV 
we set up the path
integral formulation for the noisy Burgers equation in terms of the
Martin-Siggia-Rose techniques in functional form. In section V we perform
a shift transformation of the path integral to a canonical Feynman path
integral form and discuss the canonical structure and the associated
symmetry algebra. Section VI is devoted to an asymptotic weak noise 
saddle point approximation and to the derivation of the deterministic 
coupled field equations replacing the Burgers equation. In section VII
we solve the field equations and derive nonlinear soliton and linear
diffusive mode solutions. In section VIII we discuss the dynamics
of the solitons following from the principle of least action. The dominating
morphology of a stochastically growing interface can be interpreted
in terms of a dilute soliton gas; this aspect is discussed in some detail
in section IX. The fluctuation spectrum about the soliton solutions
is basically given by the path integral, however, in section X we take
a heuristic point of view and discuss the fluctuations as 
``quantum fluctuations'' in the underlying non-Hermitian ``quantum field
theory''. Section XI is devoted to a discussion of  the scaling properties
and universality classes on the basis of the ``elementary excitations''
in the ``quantum description''. We also present a heuristic expression
for the scaling function.
Finally in section XII we present a discussion and
a conclusion.
%%%%%%%%%%%%
%%%%%%%%%%%%
%%%%%%%%%%%%
%%%%%%%%%%%%
%%%%%%%%%%%%
%%%%%%%%%%%%
%%%%%%%%%%%%
%%%%%%%%%%%%
%%%%%%%%%%%%
\section{The linear Edwards - Wilkinson equation -- the role of noise}
%%%%%%%%%%%%
%%%%%%%%%%%%
%%%%%%%%%%%%
%%%%%%%%%%%%
%%%%%%%%%%%%
%%%%%%%%%%%%
%%%%%%%%%%%%
%%%%%%%%%%%%
Here we review the properties of the linear case described by the
noise-driven Edwards-Wilkinson equation (\ref{ew1}), in particular in order
to elucidate the role of the noise.
For the slope field this equation is given by 

%{\em ew2}
%
\begin{equation}
\frac{\partial u}{\partial t} = \nu\nabla^2u+\nabla\eta 
\label{ew2}
\end{equation}
with the noise $\eta$ correlated according to Eq. (\ref{noise}), i.e., 

%{\em noise1}
%
\begin{equation}
\langle\eta(x,t)\eta(x',t')\rangle
=
\Delta\delta(x-x')\delta(t-t') ~ .
\label{noise1}
\end{equation}
The equation (\ref{ew2})
has the form of a conservation law

%{\em cl}
%
\begin{equation}
\frac{\partial u}{\partial t} = -\nabla j
\label{cl}
\end{equation}
with current

%{\em cur}
%
\begin{equation}
j = -\nu\nabla u - \eta  ~ .
\label{cur}
\end{equation}
We note that with average vanishing $\nabla u$ at the boundaries
the conservation law implies that the average off-set
in the height, $\int\nabla hdx$, is conserved.

In wave number space,
$u(k,t)=\int dx\exp{(-ikx)}u(x,t)$,
and solving Eq. (\ref{ew2} as an initial value problem averaging over the noise
according to Eq. (\ref{noise1})
we obtain for the slope correlations

%{\em tcor}
%
\begin{equation}
\langle u(k,t)u(-k,t')\rangle 
=
[\langle u(k,0)u(-k,0)\rangle_i
- \Delta/2]\exp{[-(t+t')\nu k^2]}
+\Delta/2\exp{[-|t-t'|\nu k^2]} ~ .
\label{tcor}
\end{equation}
Here $\langle\cdots\rangle_i$ denotes an average over initial values
which is assumed independent of the noise average $\langle\cdots\rangle$.
The basic
time scale is set by the wave number dependent lifetime 
$\tau(k)=1/\nu k^2$;  which diverges in the long
wave length limit $k\rightarrow 0$, characteristic of a conserved
hydrodynamical mode.
We note that
at short times compared to $\tau(k)$, which sets the time scale
for the transient
regime, $\langle u(k,t)u(-k,t')\rangle$ is non-stationary and depends 
on the initial correlations, whereas at long times $t,t'\gg\tau(k)$ 
the correlations
enter a stationary, time reversal invariant
regime and depends only on $|t-t'|$. For vanishing initial slope, $u(k,0)=0$,
we obtain in particular the mean square slope fluctuations

%{\em stcor}
%
\begin{equation}
\langle|u(k,t)|^2\rangle = \frac{\Delta}{2}
[1 - \exp{[-2t/\tau(k)]}]
\label{stcor}
\end{equation}
which approaches the saturation value $\Delta/2$ for $t\gg\tau(k)$.

More precisely, it follows from Eq. (\ref{tcor}) that for fixed $t-t'$ the
transient term can be neglected at times greater than a characteristic
crossover time $t_{co}$ of order

%{\em cot}
%
\begin{equation}
t_{co}\sim 1/(\nu k^2)\log{(1/\Delta)} ~ ,
\label{cot}
\end{equation}
depending also on the noise strength $\Delta$. This time thus defines
the onset of the stationary regime. 
For $t\gg\tau(k),t_{co}$ noise-induced fluctuations built up and 
the mean square slope fluctuations
approach the constant value $\Delta/2$. 
For $\Delta\rightarrow 0$, $t_{co}\rightarrow\infty$ and
the system never enters the stationary
regime.

The ``elementary excitation''
is the diffusive mode
$u(k,t)\propto\exp{(\pm\nu k^2t)}$. In frequency space
$u(k,\omega)=\int dt\exp{(i\omega t)}u(k,t)$ and the slope correlation function
assumes the Lorentzian diffusive form, characteristic of a hydrodynamical mode,

%{\em lor}
%
\begin{equation}
\langle u(k,\omega)u(-k,-\omega)\rangle
=
\frac{\Delta k^2}{\omega^2+(\nu k^2)^2} ~ ,
\label{lor}
\end{equation}
with diffusive poles at $\omega_k^0=\pm i\nu k^2$, a strength given by
$\Delta/\nu$ and a line width $\nu k^2$. We note that in the 
stationary regime both the decaying 
and growing modes, $u\propto\exp{(\pm\nu k^2t)}$, contribute to the stationary
correlations. Time reversal invariance is thus induced from the microscopic 
reversibility of the noise-driven system. In the transient regime for
$t\ll\tau(k)$ the initial conditions enter and we must choose the solution
propagating forward in time, $u\propto\exp{(-\nu k^2t)}$, in order to 
satisfy
causality. 

From Eq. (\ref{lor}) we also obtain the scaling function
%{\em sfew}
%
\begin{equation}
f(w)=(\Delta/2\nu)(4\pi\nu)^{-1/2}w^{-1/2}\exp{[-1/4\nu w]} ~ .
\label{sfew}
\end{equation}
in accordance with the general form in Eq. (\ref{sf}) yielding the 
EW exponents in Table 1
defining the EW universality class.
For large $w$ $f(w)\sim w^{-1/2}$; for small $w$ $f(w)\rightarrow 0$ but
with an essential singularity for $w=0$.
In frequency-wave number space the scaling form is
%{\em sfew2}
%
\begin{equation}
\langle u(k,\omega)u(-k,-\omega)\rangle=
k^{-2}g(\omega/k^2)
\label{sfew2}
\end{equation}
and we directly infer the scaling function

%{\em sfew3}
%
\begin{equation}
g(w) = \frac{\Delta}{\nu+w^2} ~ .
\label{sfew3}
\end{equation}
In Fig. 2 we have shown the slope correlation function and the 
scaling functions $f$ and $g$ in the EW case.

In contrast to the noisy Burgers equation, the EW equation does not
provide a proper description of a growing interface.
This is seen by expressing 
Eq. (\ref{ew2}) in the form

%{\em lan}
%
\begin{equation}
\frac{\partial u}{\partial t} = \nu\nabla^2\frac{\delta F}{\delta u}
+
\nabla\eta ~ ,
\label{lan}
\end{equation}
where the effective free energy is given by

%{\em free}
%
\begin{equation}
F = \frac{1}{2}\int dx u^2 ~ .
\label{free}
\end{equation}
Using the fluctuation-dissipation theorem to relate the
noise strength to an effective  temperature $T$ it then follows that
the  EW equation describes time-dependent 
fluctuations in an {\em equilibrium} 
system with
temperature $T = \Delta/2\nu$ and with an equilibrium
distribution given by the Boltzmann factor  Eq. (\ref{gauss}), i.e.,

%{\em gauss2}
%
\begin{equation}
P(u)\propto\exp{\left[-\frac{2\nu}{\Delta}F\right]} ~ .
\label{gauss2}
\end{equation}

We already here in the linear case note that  the noise 
strength $\Delta$ seems to play a special role.
Whereas $\Delta$ enters linearly in the correlation function 
$\langle uu\rangle(k,\omega)$, the limit of vanishing noise strength,
$\Delta\rightarrow 0$,  appears as an {\em essential singularity} in the
stationary distribution (\ref{gauss2}). Since the distribution $P(u)$,
appropriately generalized to the time-dependent case, is the generator
for the  correlation 
function $\langle uu\rangle$ and higher commulants, 
it is clearly
the fundamental object and the role of the noise
strength $\Delta$ as a non-perturbative parameter an important 
observation. More precisely the point is the following: 
Whereas the damping constant
$\nu$ together with the relevant wave number $k$ defines the time scale for
the transient regime where the system has memory and evolves forward in
time in an
irreversible manner, the presence of the noise  is
essential in order for the system to leave the transient regime {\em at all} 
and to enter
the stationary regime where the system is time reversal invariant,
as for example reflected in the evenness in $\omega$ in the slope
correlation function (\ref{lor}).
In the absence of the noise 
the system simply decays owing to dissipation unless it is driven by 
deterministic currents
at the boundaries. 
Imposing the noise and driving the system stochastically is
thus a {\em singular} process, as reflected
mathematically by the essential singularity in the distribution (\ref{gauss2}).

Although the above observation of the non-perturbative role of the 
noise strength $\Delta$ is a trivial statement in the linear
case where is just reflects the structure of the Boltzmann
factor, we will later show that in a more complete theory of a 
growing 
interface, described by the nonlinear noisy Burgers equation, it
is essential to
take into account non-perturbative contributions in the noise strength
$\Delta$. 
%%%%%%%%%%%%%%%%%%%%%%%%%%%%%%%%%%%
%%%%%%%%%%%%%%%%%%%%%%%%%%%%%%%%%%%
%%%%%%%%%%%%%%%%%%%%%%%%%%%%%%%%%%%
%%%%%%%%%%%%%%%%%%%%%%%%%%%%%%%%%%%
%%%%%%%%%%%%%%%%%%%%%%%%%%%%%%%%%%%
%%%%%%%%%%%%%%%%%%%%%%%%%%%%%%%%%%%
%%%%%%%%%%%%%%%%%%%%%%%%%%%%%%%%%%%
%%%%%%%%%%%%%%%%%%%%%%%%%%%%%%%%%%%
%%%%%%%%%%%%%%%%%%%%%%%%%%%%%%%%%%%
%%%%%%%%%%%%%%%%%%%%%%%%%%%%%%%%%%%
%%%%%%%%%%%%%%%%%%%%%%%%%%%%%%%%%%%
\section{The soliton mode in the noiseless Burgers equation}
%%%%%%%%%%%%%%%%%%%%%%%%%%%%%%%%%%%
%%%%%%%%%%%%%%%%%%%%%%%%%%%%%%%%%%%
%%%%%%%%%%%%%%%%%%%%%%%%%%%%%%%%%%%
%%%%%%%%%%%%%%%%%%%%%%%%%%%%%%%%%%%
%%%%%%%%%%%%%%%%%%%%%%%%%%%%%%%%%%%
%%%%%%%%%%%%%%%%%%%%%%%%%%%%%%%%%%%
%%%%%%%%%%%%%%%%%%%%%%%%%%%%%%%%%%%
%%%%%%%%%%%%%%%%%%%%%%%%%%%%%%%%%%%
%%%%%%%%%%%%%%%%%%%%%%%%%%%%%%%%%%%
%%%%%%%%%%%%%%%%%%%%%%%%%%%%%%%%%%%
It turns out that the soliton excitation in the {\em noiseless} Burgers 
equation
when properly generalized play an important role in the understanding of
the growth morphology and  strong coupling behavior of the noisy 
Burgers equation. In A 
we
discussed in some detail the soliton and diffusive mode solutions in the 
noiseless Burgers equation
and performed a linear stability analysis. Here we briefly summarize 
those 
aspects of the analysis in A which will be of importance
in the discussion of the noisy case.

The noiseless or deterministic Burgers equation has the form
\cite{burgers,saffman,jackson,whitham}

%{\em nlbur}
%
\begin{equation}
\frac{\partial u}{\partial t} = \nu\nabla^2 u + \lambda u \nabla u .
\label{nlbur}
\end{equation}
and is  a nonlinear diffusive
evolution equation with a linear  term controlled by the damping
or viscosity $\nu$ and a nonlinear mode coupling term characterized
by $\lambda$. In the context of fluid motion the nonlinear term
gives rise to convection as in the Navier Stokes equation;
for an
interface the term corresponds to a slope dependent growth.

Under time reversal $t\rightarrow -t$ and the transformation
$u\rightarrow -u$ the
equation is invariant provided $\nu\rightarrow -\nu$. This indicates
that the linear diffusive term  and the nonlinear convective or
growth term play  completely different roles. The diffusive term
is intrinsically
irreversible whereas the growth term corresponding to a mode coupling
leads to a cascade in wave number space and a generates genuine transient
growth. The transformation
$t\rightarrow -t$ is absorbed in $u\rightarrow -u$ or, alternatively,
$\lambda\rightarrow -\lambda$, corresponding to a change of growth direction.
We also note that the equation is invariant under the parity
transformation $x\rightarrow -x$ provided $u\rightarrow -u$.
This feature is related
to the presence of a single spatial derivative in the growth term
and implies that the equation only supports solitons or shocks with one
parity, that is {\em right hand} solitons.
Finally, the Burgers equation is invariant under the Galilean symmetry
group (\ref{galx}-\ref{galu}), that is a Galilean boost to a frame moving
with velocity $\lambda u_0$ is absorbed
by a shift in the slope field by $u_0$.

The irreversible and diffusive structure of Eq. (3.1) implies that an
initial disturbance eventually decays owing to the damping term $\nu\nabla^2 u$.
In the linear case the slope field decays
by simple diffusion $u(x,t)\propto\exp{(-\nu k^2t)}\exp{(\pm ikx)}$
as discussed in section II. In the presence of the nonlinear
mode coupling term the equation also supports localized soliton or 
kink profiles
\cite{scott,bishop,bernasconi,fogedby5}
with given parity. In the static case
the symmetric positive parity or {\em right hand} soliton has the form

%{\em ssol, wn}
%
\begin{eqnarray}
u(x) =&& u_+\tanh{[k_s(x-x_0)]}
\label{ssol}
\\
k_s=&&\lambda u_+/2\nu ~ .
\label{wn}
\end{eqnarray}
We have introduced the characteristic wave number $k_s$ setting the
inverse length scale associated with the static soliton, 
$x_0$ denotes the center of mass
position. The width of the soliton is of order $1/k_s$ and 
depends on the amplitude $u_+$. In the inviscid
limit $\nu\rightarrow 0$ or for strong coupling $\lambda\rightarrow\infty$,
the wave number $k_s\rightarrow\infty$ and
the soliton reduces to a sharp shock wave discontinuity. 

Boosting the static soliton in Eq. (\ref{ssol}) to a finite propagation 
velocity $v$ and
at the same time shifting $u$ we obtain, denoting the right and left
boundary values by $u_\pm$, the soliton solution

%{\em msol}
%
\begin{equation}
u(x,t) = \frac{u_++u_-}{2} + \frac{u_+-u_-}{2}
\tanh{[\frac{\lambda}{4\nu}(u_+-u_-)(x-vt-x_0)]}
\label{msol}
\end{equation}
with velocity $v$ given by the soliton condition

%{\em solcon}
%
\begin{equation}
u_+ + u_- = -\frac{2v}{\lambda} ~ .
\label{solcon}
\end{equation}
We note that the soliton condition (\ref{solcon}) is consistent 
with the fundamental
nonlinear Galilean invariance and  remains invariant under
the transformation: $v\rightarrow v+\lambda u_0$ and 
$u_\pm\rightarrow u_\pm -u_0$.
Also, unlike the case for the Lorentz invariant $\phi^4$ and 
sine-Gordon evolution equations \cite{scott},
the propagation velocity in the present case is tied to the
amplitude boundary values of the soliton - a feature of the nonlinear
Galilean invariance.
In Fig. 3 we have depicted the {\em right hand} soliton solution given by
Eq. (\ref{msol}) and the associated height field $h$.

In the linear case for $\lambda =0$ case the diffusive modes with dispersion 

%{\em disp}
%
\begin{equation}
\omega_k^0 = -i\nu k^2
\label{disp}
\end{equation}
``exhaust'' the spectrum of relaxational modes. For $\lambda\neq 0$ 
the soliton profile acts as a reflectionless Bargman potential
giving rise to a bound state at zero frequency, corresponding to
the translation mode of the soliton - the Goldstone mode restoring the
broken translational invariance, and a band of phase-shifted diffusive 
scattering modes \cite{fogedby5}. The resulting change of density of 
states is in accordance
with Levinson's theorem in that the potential traps a bound state and depletes
the continuum of one state.
In the presence of the soliton the diffusive modes furthermore develop 
a gap in the spectrum of $\omega_k$ as depicted in Fig. 4,

%{\em dispgap}
%
\begin{equation}
\omega_k = -i\nu(k^2+k_s^2) ~ .
\label{dispgap}
\end{equation}

An asymptotic analysis of the noiseless Burgers equation in the
inviscid limit $\nu\rightarrow 0$ \cite{aurell} shows that an 
initial configuration
breaks up into a ``gas'' of propagating and coalescing kinks connected by ramp
solutions of the form $u\propto \mbox{const} - x/\lambda t$. This allows for
the following
qualitative picture of the transient time evolution:
Although the nonlinear mode coupling term is incompatible
with a proper  superposition principle we can still along the lines of
the evolution of integrable one dimensional evolution equations 
\cite{scott}
envisage
that an 
initial configuration ``contains'' a number of  {\em right hand} 
solitons connected by ramps. In the course of time the solitons
propagate and coalesce.
Superposed on the soliton gas is a gas of phase-shifted
diffusive modes. As discussed in A the gap in the diffusive
spectrum can be associated with the current flowing towards the
center of the solitons. The damping of the configuration predominantly
takes place at the center of the soliton where $u$ varies rapidly thus
enhancing 
the damping term $\nu\nabla^2 u$. We also note
that only parity breaking {\em right hand} solitons are generated in the
noiseless Burgers equation.
In Fig. 5 we have shown the transient evolution of the slope field
and the associated height field.
%%%%%%%%%%%%%%%%%%%%%%%%%%%%%%%%%%%%%%%
%%%%%%%%%%%%%%%%%%%%%%%%%%%%%%%%%%%%%%%
%%%%%%%%%%%%%%%%%%%%%%%%%%%%%%%%%%%%%%%
%%%%%%%%%%%%%%%%%%%%%%%%%%%%%%%%%%%%%%%
%%%%%%%%%%%%%%%%%%%%%%%%%%%%%%%%%%%%%%%
%%%%%%%%%%%%%%%%%%%%%%%%%%%%%%%%%%%%%%%
%%%%%%%%%%%%%%%%%%%%%%%%%%%%%%%%%%%%%%%
%%%%%%%%%%%%%%%%%%%%%%%%%%%%%%%%%%%%%%%
%%%%%%%%%%%%%%%%%%%%%%%%%%%%%%%%%%%%%%%
\section{Path integral representation of the noisy Burgers equation}
%%%%%%%%%%%%%%%%%%%%%%%%%%%%%%%%%%%%%%%
%%%%%%%%%%%%%%%%%%%%%%%%%%%%%%%%%%%%%%%
%%%%%%%%%%%%%%%%%%%%%%%%%%%%%%%%%%%%%%%
%%%%%%%%%%%%%%%%%%%%%%%%%%%%%%%%%%%%%%%
%%%%%%%%%%%%%%%%%%%%%%%%%%%%%%%%%%%%%%%
%%%%%%%%%%%%%%%%%%%%%%%%%%%%%%%%%%%%%%%
%%%%%%%%%%%%%%%%%%%%%%%%%%%%%%%%%%%%%%%
%%%%%%%%%%%%%%%%%%%%%%%%%%%%%%%%%%%%%%%
In this section we begin the analysis of the noisy Burgers equation.
In our discussion in section II of the linear EW equation we noticed
that the noise 
strength $\Delta$
enters in a non-perturbative way in the stationary distribution in
Eq. (\ref{gauss}). Whereas this, of course, is a trivial observation
in the linear case
since the EW equation describes fluctuations in equilibrium and 
$\Delta\propto T$, that is the singularity structure is the same as the 
low $T$ limit of the Boltzmann factor $\exp{(-E/T)}$,
the presence of the nonlinear mode coupling growth term in the noisy
Burgers equation  renders the situation much more subtle.
We are now dealing with
an intrinsically nonequilibrium situation. The noise drives the system
into a far-from-equilibrium stationary state and equilibrium statistical 
mechanics does not apply. 
On the other hand, from our study of the noiseless Burgers equation,
we have learned that the soliton excitations play an 
important role in the  
dynamics of the morphology of a growing interface  and is a direct 
signature of the nonlinearity. The issue facing us is then
how to include both the non-perturbtive aspects of the noise and 
the nonlinear soliton structure  in a consistent
way.
It turns out that the functional formulation of the Martin-Siggia-Rose
techniques provides the appropriate formal and practical language
for such an approach \cite{msr,dedom1,dedom2,dedom3,janssen2,baussch}.

Our starting point is the
noisy Burgers equation (\ref{bur}) for the fluctuating slope field
$u$, i.e., 

%{\em bur2}
%
\begin{equation}
\frac{\partial u}{\partial t} = \nu\nabla^2 u + \lambda u \nabla u 
+\nabla\eta
\label{bur2}
\end{equation}
which has the structure of  conserved nonlinear Langevin equation with 
current

%{\em cur2}
%
\begin{equation}
j=-\nabla u - \frac{\lambda}{2}u^2 -\eta ~ .
\label{cur2}
\end{equation}
For the
noise we assume a Gaussian distribution
%

%{\em ng}
\begin{equation}
P(\eta)\propto\exp{\left[-\frac{1}{2\Delta}\int dxdt\eta(xt)^2\right]} ~ ,
\label{ng}
\end{equation}
where $\eta$ is correlated according to Eq. (\ref{noise}), i.e., 
%

%{\em noise2}
\begin{equation}
\langle\eta(x,t)\eta(x',t')\rangle
=
\Delta\delta(x-x')\delta(t-t') ~ .
\label{noise2}
\end{equation}
%.

Unlike the transient relaxation of an initial value configuration
described by the deterministic Burgers equation,
the noisy Burgers equation is driven continuously
by the conserved noise $\nabla\eta$, corresponding to a fluctuating
component of the current $j$ in Eq. (\ref{cur2}). 
Energy is fed into the system via the noise and dissipated by the 
linear damping term. The nonlinear mode coupling gives rise to a 
cascade in wave number space corresponding to ``dissipative structures''
in the growth morphology. This mechanism changes the probability 
distributions and associated correlations (moments), scaling exponents,
and scaling functions from the EW case in section II. In other words,
the 
equation (\ref{bur2}) acts as a nonlinear box which transforms the
input noise $\nabla\eta$ to an output slope field $u$.

In the Martin-Siggia-Rose techniques the probability distribution 
for the slope field $P(u)$ and the correlations
$\langle uu\rangle_\eta$ are conveniently derived from an effective partition
function or generator \cite{zinn-justin}

%{\em gen}
%
\begin{equation}
Z(\mu) = 
\left\langle\exp{\left[i\int dxdtu(x,t)\mu(x,t)\right]}\right\rangle_\eta ~ .
\label{gen}
\end{equation}
Here $\mu(x,t)$ is a  generalized chemical potential
or external conjugate field  coupling to the
slope field $u(x,t)$ and $\langle\cdots\rangle_\eta$ denotes an 
average over the input noise $\eta$, implementing the nonlinear
stochastic relationship provided by the Burgers equation (\ref{bur2}). 
In terms of $Z$ we have for example the probability distribution
$P(u)=\langle\delta(u-u(x,t))\rangle_\eta$,

%{\em pd}
%
\begin{equation}
P(u(x,t)) = \int\prod_{xt}d\mu\exp{\left[-i\int dxdtu(x,t)\mu(x,t)\right]}
Z(\mu(x,t))
\label{pd}
\end{equation}
and the correlation function

%{\em cor}
%
\begin{equation}
\langle u(x,t)u(x',t')\rangle
= -\left[\frac{\delta Z(\mu)}{\delta\mu(x,t)\delta\mu(x',t')}\right]_{\mu=0} ~ ;
\label{cor}
\end{equation}
higher moments are derived in a likewise manner.
In order to incorporate the nonlinear constraint imposed by the
Burgers equation we insert the identity 

%{\em id}
%
\begin{equation}
\int\prod_{xt}du
\delta\left(\frac{\partial u}
{\partial t}-\nu\nabla^2u-\lambda u\nabla u-\nabla\eta\right)
=1
\label{id}
\end{equation}
in the partition function $Z(\mu)$; for a first order evolution equation
one can show that causality implies that the Jacobian relating $du$ to 
$\partial u/\partial t$ equals unity \cite{footnote2}. Finally, 
exponentiating the delta function
constraint in Eq. (\ref{id}) and averaging over the noise distribution 
according to Eq. (\ref{ng}) we obtain 

%{\em gen2}
%
\begin{equation}
Z(\mu) = \int\prod_{xt}dudp\exp{[iG]}\exp{\left[i\int dxdtu\mu\right]} ~ ,
\label{gen2}
\end{equation}
where the effective functional $G$ is given by

%{\em f}
%
\begin{equation}
G = \int dxdt\left[p(\frac{\partial u}
{\partial t} -\nu\nabla^2 u-\lambda u\nabla u)
+\frac{i}{2}\Delta(\nabla p)^2\right] ~.
\label{f}
\end{equation}

The path or functional integral (\ref{gen2}) with $G$ given by Eq. (\ref{f}) 
effectively replaces the stochastic Burgers equation (\ref{bur2}). 
The path integral is
deterministic and the noise $\eta$ is replaced by the different configurations
or paths contributing to $Z$. In this sense $Z$ is an effective 
partition function for the dynamical problem and $G$ an effective 
Hamiltonian, analogous to the Hamiltonian in the partition function 
$Z=\sum\exp{(-H/T)}$
in equilibrium statistical mechanics.
We also note that the transcription of the Burgers equation to a path 
integral leads to the appearance of an additional noise field
$p$, arising from the exponentiation of the delta function constraint in 
Eq. (\ref{id})\cite{msr,dedom1,dedom2,dedom3,janssen2,baussch},
and replacing the stochastic noise in Eq. (\ref{bur2}).

Since the path integral formulation provides a field theoretical
framework allowing for functional and diagrammatic techniques,
Feynman rules, skeleton graphs, Ward identities, etc., it is mostly used 
in order to generate perturbation expansions
in powers of the 
the nonlinear coupling  $\lambda u\nabla u$
\cite{frey1,tauber,frey2,zinn-justin,som1,som2,som3}. 
It is, however, worthwhile noting that such a field theoretic
expansion has
precisely the same structure as the one produced by directly
iterating the Burgers equation (\ref{bur2}) in powers of $\lambda u\nabla u$
and averaging over the noise term by term according to Eq. (\ref{noise2}).

Also, corroborating our remarks in section II, we notice from the
structure of the path integral (\ref{gen2}-\ref{f}) that the noise strength
$\Delta$ appears as a singular parameter in the sense that 
$\Delta\rightarrow 0$ gives rise to the singular delta function
constraint for the Burgers equation. This limit is, however,
much more transparent when we express $Z$ in a ``canonical form''.
%%%%%%%%%%%%%%%%%%%%%%%%%%%%%%%%%%
%%%%%%%%%%%%%%%%%%%%%%%%%%%%%%%%%%
%%%%%%%%%%%%%%%%%%%%%%%%%%%%%%%%%%
%%%%%%%%%%%%%%%%%%%%%%%%%%%%%%%%%%
%%%%%%%%%%%%%%%%%%%%%%%%%%%%%%%%%%
\section{Canonical transformation to a Hamiltonian form - Phase space
path integral - Symmetries}
%%%%%%%%%%%%%%%%%%%%%%%%%%%%%%%%%%
%%%%%%%%%%%%%%%%%%%%%%%%%%%%%%%%%%
%%%%%%%%%%%%%%%%%%%%%%%%%%%%%%%%%%
%%%%%%%%%%%%%%%%%%%%%%%%%%%%%%%%%%
%%%%%%%%%%%%%%%%%%%%%%%%%%%%%%%%%%
%%%%%%%%%%%%%%%%%%%%%%%%%%%%%%%%%%
%%%%%%%%%%%%%%%%%%%%%%%%%%%%%%%%%%
By inspection of the path integral in Eqs. (\ref{gen2}-\ref{f}) 
we notice that it
has the same structure as the usual phase space Feynman path integral 
\cite{zinn-justin,feynman,das} as regards the kinetic term 
$p\partial u/\partial t$ in $F$ but that otherwise $p$ and $u$ do not
appear in a canonical combination. This situation can,
however, be remedied by
performing a simple complex shift of the noise variable $p$ 

%{\em sh}
%
\begin{equation}
p = \frac{\nu}{\Delta}(iu - \varphi)
\label{sh}
\end{equation}
in Eqs. (\ref{gen2}-\ref{f}). Assuming that the path integral 
operates in a space
time LT box, i.e., $|x|<L/2$ and $|t|<T/2$,
and imposing periodic or vanishing boundary conditions for $u$ and $\varphi$
in order to eliminate total derivatives,
the partition function 
$Z(\mu)$ can be expressed as

%{\em gen3}
%
\begin{equation}
Z(\mu)= \mbox{const}\int\prod_{xt} dud\varphi
\exp{\left[i\frac{\nu}{\Delta}S\right]}
\exp{\left[i\int dxdt u\mu\right]} ~ ,
\label{gen3}
\end{equation}
where the action $S$ is given by the canonical form
\cite{landau1}

%{\em ac}
%
\begin{equation}
S = 
\int dxdt\left[u\frac{\partial\varphi}{\partial t} - {\cal H}(u,\varphi)\right]
\label{ac}
\end{equation}
with the complex Hamiltonian density

%{\em ham}
%
\begin{equation}
{\cal H} = -i\frac{\nu}{2}[(\nabla u)^2 + (\nabla\varphi)^2]
+ \frac{\lambda}{2} u^2\nabla\varphi ~ .
\label{ham}
\end{equation}
The Hamiltonian density consists of two terms: 
A relaxational or
irreversible harmonic component, $-i(\nu/2)[(\nabla u)^2+(\nabla\varphi)^2]$,
corresponding to the diffusive aspects of a growing interface,
i.e., the linear damping, and a
nonlinear reversible mode coupling component, $(\lambda/2)u^2\nabla\varphi$,
associated
with the drive $\lambda$.

One feature of the transcription of the noisy Burgers
equation to a canonical path integral form is that the 
effective Hamiltonian
density (\ref{ham}) driving the dynamics of the system is in general complex.
This particular aspect
was also encountered in the treatment
in L where the growth term in the spin chain Hamiltonian
turned out to be complex. 
We also notice that the doubling of variables, i.e., the replacement of
the stochastic noise $\eta$ by an additional noise field $\varphi$ in the
path integral, was also encountered in the treatment in L in the canonical
oscillator representation of the spin variables.

It is here instructive to compare the above path integral for the 
relaxational
growth dynamics of the Burgers equation with the usual phase space path
integral formulation in quantum mechanics or quantum field theory
\cite{feynman,das}.
Here the partition function has the form

%{\em qgen}
%
\begin{equation}
Z = \int\prod_{xt}dpdq\exp{\left[\frac{i}{\hbar}S\right]}
\label{qgen}
\end{equation}
with the classical action

%{\em qac}
%
\begin{equation}
S = \int dxdt\left[p\frac{\partial q}{\partial t} - {\cal H}(p,q)\right] ~ ,
\label{qac}
\end{equation}
where $p$ and $q$ are considered canonically conjugate variables 
and ${\cal H}(p,q)$ the
usual classical Hamiltonian density.

Comparing Eqs. (\ref{qgen}-\ref{qac}) with Eqs. (\ref{gen3}-\ref{ac}) 
it is evident that  
the structures 
of the two path integral formulations are 
quite similar and we are led to identify the noise strength
$\Delta/\nu$ with an ``effective'' Planck constant and the Hamiltonian
density ${\cal H}$ as the generator of the dynamics.
The classical limit
thus corresponds to the weak noise limit $\Delta\rightarrow 0$ and in
analogy with the quasi-classical or WKB approximation in quantum mechanics,
$\Delta\rightarrow 0$, constitutes a singular limit in accordance with
our previous remarks. The partition function $Z$ with the action $S$ given 
by Eqs. (\ref{gen3}-\ref{ham}) thus constitutes the required generalization of 
the stationary distribution
$P(u)\propto\exp{[-(\nu/\Delta)\int dx u^2]}$ in Eq. (\ref{gauss}) to the
time-dependent case \cite{footnote3}.
By comparison we furthermore conclude that the slope field $u$ and the
noise field $\varphi$, replacing the Gaussian noise in Eq. (\ref{bur2}), are 
canonically conjugate momentum and coordinate variables 
satisfying the Poisson bracket algebra
\cite{footnote4}

%{\em pa}
%
\begin{equation}
\{u(x),\varphi(x')\} = \delta(x-x') 
\label{pa}
\end{equation}
and that the Hamiltonian  or energy 

%{\em ha}
%
\begin{equation}
H = \int dx{\cal H}=
\int dx\left[-i\frac{\nu}{2}\left[(\nabla u)^2 + (\nabla\varphi)^2\right]
+\frac{\lambda}{2}u^2\nabla\varphi\right] 
\label{ha}
\end{equation}
is the generator of time translations according to the equations of motion
\begin{eqnarray}
\frac{\partial u}{\partial t}&&= \{H,u\}
\label{em1}
\\
\frac{\partial\varphi}{\partial t}&&= \{H,\varphi\} ~ .
\label{em2}
\end{eqnarray}
Drawing
on the mechanical analogue the momentum $P$, the generator of translations in
space, is also easily identified from the basic transformation properties,

%{\em mom1, mom2}
%
\begin{eqnarray}
\nabla u =&& \{P,u\}
\label{mom1}
\\
\nabla\varphi =&& \{P,\varphi\} ~ ,
\label{mom2}
\end{eqnarray}
and it follows that

%{\em mom, md}
%
\begin{eqnarray}
P =&& \int dx g \\
\label{mom}
g =&& u\nabla\varphi ~ ,
\label{md}
\end{eqnarray}
where $g$ is the momentum density.

In order to elucidate the canonical structure of the path integral 
(\ref{gen3}-\ref{ham}) and
the analogy with the usual phase space Feynman path integral we have 
generated a complex Hamiltonian (\ref{ham}). Note, however, that by formally
rotating the noise field in phase space 
$\varphi\rightarrow i\varphi$ the Hamiltonian and the 
action become purely imaginary leading to a real path integral.

The symmetries discussed in the context of the quantum spin chain
representation in L  are also easily recovered here. Noting
that ${\cal H}$ is invariant under a constant shift of the noise field,
$\varphi\rightarrow\varphi + \varphi_0$, we infer that the
integrated slope field

%{\em is}
%
\begin{equation}
M = \int dx u
\label{is}
\end{equation}
i.e., the total off-set of the
height field, $h = \int dx u$, across the interface, is a constant of
motion,

%{\em com1}
%
\begin{equation}
\{H,M\} = 0 ~ .
\label{com1}
\end{equation}
This is consistent with the local conservation law
following from the structure
of the Burgers equation, but is here a consequence of the
structure of the path integral. The invariance under a shift
of $\varphi$ is equivalent to the invariance of the Burgers equation
under a shift
of the noise $\eta$ in the noise term $\nabla\eta$. Similarly, under a
constant shift of the slope field $u\rightarrow u + u_0$, we have,
introducing the momentum density $g$, 
${\cal H}\rightarrow{\cal H}+\lambda u_0g +(\lambda/2)u_0^2\nabla g$, or
since the last term is a total derivative, $H\rightarrow H +\lambda u_0 P$,
corresponding to an associated Galilean transformation with velocity
$-\lambda u_0$. For the integrated noise field 

%{\em in}
%
\begin{equation}
\Phi = \int dx\varphi
\label{in}
\end{equation}
we thus obtain the Poisson bracket algebra

%{\em com2}
%
\begin{equation}
\{H,\Phi\} = \lambda P
\label{com2}
\end{equation}
which together with

%{\em com3, com4}
\begin{eqnarray}
\{H,M\} = && \{H,P\} = 0
\label{com3}
\\
\{P,\Phi\} = && \{P,M\} = 0
\label{com4}
\end{eqnarray}
and

%{\em com5}
%
\begin{equation}
\{\Phi,M\} = L
\label{com5}
\end{equation}
defines the symmetry algebra.
We note again that the nonlinear coupling strength $\lambda$ enters the Poisson
bracket (\ref{com2}) and thus is  a structural constant of the symmetry group.

We finally wish to comment on the properties of the path integral in 
Eqs. (\ref{gen3}-\ref{ham}) under time reversal $t\rightarrow -t$.
By construction the path integral applies at late times
compared to any initial time $t_0$, defining the initial value of the
slope configuration $u_0$. This implies that the noise in the Burgers
equation  has driven the system into a stationary time regime
and that the transients associated with $u_0$ have died out. 
In the linear case for
$\lambda = 0$, we note by inspection
that the path integral is invariant under the combined
operation $t\rightarrow -t$ and $\varphi\rightarrow -\varphi$, implying that
the slope correlations are not only stationary but also invariant under
time reversal. This is in agreement with the  analysis of the 
noisy
EW equation in section II where we obtained 
$\langle uu\rangle(k,\omega)=\Delta k^2/[\omega^2+(\nu k^2)^2]$,
implying that $\langle uu\rangle(x,t)$ depends on $|t|$.
This is consistent with the description of an equilibrium interface 
and is just
an expression of
microscopic
reversibility. In the presence of the drive for $\lambda\neq 0$
the path integral is invariant under the combined transformation
$t\rightarrow -t$, $\varphi\rightarrow -\varphi$, and 
$\lambda\rightarrow -\lambda$
showing that the term $(\lambda/2)u^2\nabla\varphi$ in the Hamiltonian
(\ref{ha}) gives rise to a proper growth direction  thereby breaking time
reversal invariance, that is we are dealing with a genuine nonequilibrium
phenomena.
%%%%%%%%%%%%%%%%%%%%%%%%%%%%%%%%%%%%%%%%%
%%%%%%%%%%%%%%%%%%%%%%%%%%%%%%%%%%%%%%%%%
%%%%%%%%%%%%%%%%%%%%%%%%%%%%%%%%%%%%%%%%%
%%%%%%%%%%%%%%%%%%%%%%%%%%%%%%%%%%%%%%%%%
%%%%%%%%%%%%%%%%%%%%%%%%%%%%%%%%%%%%%%%%%
%%%%%%%%%%%%%%%%%%%%%%%%%%%%%%%%%%%%%%%%%
%%%%%%%%%%%%%%%%%%%%%%%%%%%%%%%%%%%%%%%%%
\section{Field equations in the weak noise limit - 
Saddle point approximation}
%%%%%%%%%%%%%%%%%%%%%%%%%%%%%%%%%%%%%%%%%
%%%%%%%%%%%%%%%%%%%%%%%%%%%%%%%%%%%%%%%%%
%%%%%%%%%%%%%%%%%%%%%%%%%%%%%%%%%%%%%%%%%
%%%%%%%%%%%%%%%%%%%%%%%%%%%%%%%%%%%%%%%%%
%%%%%%%%%%%%%%%%%%%%%%%%%%%%%%%%%%%%%%%%%
%%%%%%%%%%%%%%%%%%%%%%%%%%%%%%%%%%%%%%%%%
%%%%%%%%%%%%%%%%%%%%%%%%%%%%%%%%%%%%%%%%%
%%%%%%%%%%%%%%%%%%%%%%%%%%%%%%%%%%%%%%%%%
%%%%%%%%%%%%%%%%%%%%%%%%%%%%%%%%%%%%%%%%%
The basic structure of the path integral (\ref{gen3}-\ref{ham}) is 
illustrated by the simple 
one dimensional integral, 

%{\em int}
%
\begin{equation}
I(\Delta) = \int du\exp{\left[i\frac{1}{\Delta}S(u)\right]}\exp{[i\mu u]} ~ ,
\label{int}
\end{equation}
where $\Delta$ is the small parameter (the noise strength). In the limit
$\Delta\rightarrow 0$ the integral $I(\Delta)$ is approximated by a 
steepest descent
calculation which amounts to an expansion of $S(u)$ about an extremum
$u_0$, $S(u)\sim S(u_0)+\frac{1}{2}S''(u_0)(u-u_0)^2$
and a subsequent calculation of a Gaussian integral. For small
$\Delta$ we then obtain

%{\em int2}
%
\begin{equation}
I(\Delta) = \exp{[i\frac{1}{\Delta}S(u_0)]}\exp{[i\mu u_0]}
\exp{\left[-i\mu^2\frac{\Delta}{2S''(u_0)}\right]}
\left[\frac{-2\pi i\Delta}{S''(u_0)}\right]^{1/2} ~ .
\label{int2}
\end{equation}
The leading contribution to $I(\Delta)$ is given by $\exp{[iS(u_0)/\Delta]}$
and is thus 
determined by the extremal value of the action. This part,
however, goes along with a multiplicative factor, 
$[-2\pi i\Delta/S''(u_0)]^{1/2}$, arising from the Gaussian integral 
sampling the
fluctuations about the stationary points; this term is the first in an 
asymptotic expansion in powers of $\Delta^{1/2}$. We notice that there is
an essential singularity for $\Delta =0$, signalling the non-perturbative
aspects of a steepest descent calculation; the result cannot be obtained
as a perturbation expansion in powers of $\Delta$. The analysis of the 
path integral now essentially follows the 
same procedure but is
rendered much more difficult owing to the field theoretical
phase space structure of the problem.
In Fig. 6 we have depicted the principle of a saddle point or
steepest descent calculation of $I(\Delta)$.

In the weak noise limit $\Delta\rightarrow 0$ the asymptotically leading
contribution to the path integral  thus arises from configurations
or paths $(u,\varphi)$ corresponding to an extremum or stationary point
of the action $S$. Invoking the variational condition
$\delta S = 0$ with respect to  independent variations of the slope
field $u$ and the canonically conjugate noise field $\varphi$,
$\delta u$ and $\delta\varphi$, with vanishing variations 
at the boundaries of the space time LT box,
we readily infer, using Eq. (\ref{ha}), the classical 
equations of motion 
\cite{landau1}

%{\em eqa ,eqb}
%
\begin{eqnarray}
\frac{\partial u}{\partial t}&&
= -\frac{\delta H}{\delta\varphi} 
= {\{H,u\}} 
\label{eqa}
\\
\frac{\partial\varphi}{\partial t}&&
= +\frac{\delta H}{\delta u} 
= {\{H,\varphi\}} ~ .
\label{eqb}
\end{eqnarray}
Implementing the functional derivation or, equivalently, using the
Poisson bracket relation (\ref{pa}), we obtain

%{\em eq1, eq2}
%
\begin{eqnarray}
\frac{\partial u}{\partial t}&&
=-i\nu\nabla^2\varphi + \lambda u\nabla u 
\label{eq1}
\\
\frac{\partial\varphi}{\partial t} &&
=+i\nu\nabla^2 u  + \lambda u\nabla \varphi ~ .
\label{eq2}
\end{eqnarray}

The above coupled field equations (\ref{eq1}-\ref{eq2}) are a fundamental 
result of the
present analysis. They provide a {\em deterministic} description of the noisy
Burgers equation in the asymptotic non-perturbative 
weak noise limit.
The equations have the same form as the ones derived in L  based
on the quasi-classical limit of the quantum spin chain representation.
Furthermore, the parameter identification is in accordance with the
``quantum representation'' of the Fokker-Planck equation.
As regards the considerations in L this 
demonstrates that the precise
identification of the quasi-classical limit is in fact a weak noise limit
in the exact path integral representation of the Burgers equation.

First of all we observe that the field equation for the slope field $u$ has 
the form
of a conservation law, $\partial u/\partial t = -\nabla u$, with current
$j = -(\lambda/2)u^2+i\nu\nabla\varphi$. The fluctuating component
in the current in the
noisy Burgers equation, $j = i(\lambda/2)u^2 - \nu\nabla u -\eta$, is thus
replaced by the noise field $\varphi$ and admissible solutions
must yield an imaginary noise field in order to render a real current,
corroborating our remarks in the previous section. The field equation for
the noise field is parametrically coupled to the slope field and in
the presence of the coupling $\lambda$ driven by the momentum density
$g=u\nabla\varphi$.

Secondly, we confirm that the field equations are 
invariant under the nonlinear
Galilean transformation (\ref{galx}-\ref{galu}),

%{\em gal1, gal2}
%
\begin{eqnarray}
u(x,t)&&\rightarrow u(x-\lambda u_0t,t) - u_0 
\label{gal1}
\\
\varphi(x,t)&&\rightarrow\varphi(x-\lambda u_0t,t) 
\label{gal2}
\end{eqnarray}
and under an arbitrary shift in $\varphi$

%{\em fishift}
%
\begin{equation}
\varphi(x,t)\rightarrow \varphi(x,t) - \varphi_0  ~ .
\label{fishift}
\end{equation}
in accordance with the general discussion of the symmetry algebra 
and consistent with the symmetry properties of the noiseless and noisy
Burgers equations.

One final comment on the classical zero noise limit. We maintain that
in the asymptotic non-perturbative weak noise limit the coupled field 
equations provide the correct description
of the leading behavior of the noisy Burgers equation. In order to obtain
the noiseless Burgers equation discussed in section III we must confine the
noise field strictly to the line $\varphi = iu$ in 
$(u,\varphi)$ phase space in which 
case both field equations  reduce to the noiseless 
Burgers equation (\ref{nlbur}).
Note, however, that setting $\varphi = -iu$ we
obtain the noiseless Burgers equation with $\nu$ replaced by $-\nu$ supporting
{\em growing} linear modes and a {\em left hand} nonlinear soliton mode - the
missing modes necessary in order to describe the correct morphology
in the  noise-driven stationary regime.
In other words, we anticipate that the lines $\varphi=\pm iu$
define regions for the stationary steepest descent or 
saddle point solutions of the
field equations. The vicinity of these lines correspond to 
the Gaussian fluctuations about the stationary points, that is the linear
diffusive
modes. This picture will in fact be will be borne out
when we turn to an analysis of the field equations in the
next section. In Fig. 7 we have shown the extremal paths, 
corresponding to the saddle point solutions and the nearby paths characterizing
the fluctuations in
$(u,\varphi)$ phase space.
%%%%%%%%%%%%%%%%%%%%%%%%%%%%%%%%
%%%%%%%%%%%%%%%%%%%%%%%%%%%%%%%%
%%%%%%%%%%%%%%%%%%%%%%%%%%%%%%%%
%%%%%%%%%%%%%%%%%%%%%%%%%%%%%%%%
%%%%%%%%%%%%%%%%%%%%%%%%%%%%%%%%
%%%%%%%%%%%%%%%%%%%%%%%%%%%%%%%%
%%%%%%%%%%%%%%%%%%%%%%%%%%%%%%%%
%%%%%%%%%%%%%%%%%%%%%%%%%%%%%%%%
%%%%%%%%%%%%%%%%%%%%%%%%%%%%%%%%
\section{Soliton and diffusive mode solutions of the field equations}
%%%%%%%%%%%%%%%%%%%%%%%%%%%%%%%%
%%%%%%%%%%%%%%%%%%%%%%%%%%%%%%%%
%%%%%%%%%%%%%%%%%%%%%%%%%%%%%%%%
%%%%%%%%%%%%%%%%%%%%%%%%%%%%%%%%
%%%%%%%%%%%%%%%%%%%%%%%%%%%%%%%%
%%%%%%%%%%%%%%%%%%%%%%%%%%%%%%%%
%%%%%%%%%%%%%%%%%%%%%%%%%%%%%%%%
%%%%%%%%%%%%%%%%%%%%%%%%%%%%%%%%
%%%%%%%%%%%%%%%%%%%%%%%%%%%%%%%%
%%%%%%%%%%%%%%%%%%%%%%%%%%%%%%%%
%%%%%%%%%%%%%%%%%%%%%%%%%%%%%%%%
The field equations (\ref{eq1}-\ref{eq2}) for
$u$ and $\varphi$
constitute a set of nonlinear coupled partial differential equations. The
general solution is not known. Unlike the noiseless Burgers equation
which can be solved 
by means of the nonlinear 
Cole-Hopf transformation, similar substitutions do not seem to work for
the field equations. Presently, it is not known whether the field equations
belong to the small class of nonlinear 
evolution equations
which can be integrated partly or completely by means of the inverse 
scattering method and related techniques.
We are therefore obliged to choose a more pedestrian approach and search for
special solutions to the equations \cite{scott,bishop,fogedby5}.
%%%%%%%%%%%%%%%%%%%%%%%%%%%%%%%
%%%%%%%%%%%%%%%%%%%%%%%%%%%%%%%%
%%%%%%%%%%%%%%%%%%%%%%%%%%%%%%%%
%%%%%%%%%%%%%%%%%%%%%%%%%%%%%%%%
%%%%%%%%%%%%%%%%%%%%%%%%%%%%%%%%
%%%%%%%%%%%%%%%%%%%%%%%%%%%%%%%%
%%%%%%%%%%%%%%%%%%%%%%%%%%%%%%%%
\subsection{Stationary states}
%%%%%%%%%%%%%%%%%%%%%%%%%%%%%%%%
%%%%%%%%%%%%%%%%%%%%%%%%%%%%%%%%
%%%%%%%%%%%%%%%%%%%%%%%%%%%%%%%%
%%%%%%%%%%%%%%%%%%%%%%%%%%%%%%%%
%%%%%%%%%%%%%%%%%%%%%%%%%%%%%%%%
%%%%%%%%%%%%%%%%%%%%%%%%%%%%%%%%
%%%%%%%%%%%%%%%%%%%%%%%%%%%%%%%%
%%%%%%%%%%%%%%%%%%%%%%%%%%%%%%%%
%%%%%%%%%%%%%%%%%%%%%%%%%%%%%%%%
We note that the constant slope-constant noise configurations

%{\em st1, st2}
%
\begin{eqnarray}
u=&&u_0
\label{st1}
\\
\varphi=&&\varphi_0
\label{st2}
\end{eqnarray}
are trivial saddle point solutions with vanishing energy, momentum,
and action. They form an infinitely degenerate set and correspond to the 
zero-energy aligned ferromagnetic spin states discussed in L.
The degenerate
stationary slope configurations are related by a Galilean transformation
and we shall in general choose a state with vanishing slope corresponding
to a horizontal interface. As regards the noise field we are free to
choose it equal to zero. In the phase space plot in Fig. 7 
the background stationary state or the ``vacuum'' thus corresponds to
the origin $(u,\varphi) = (0,0)$.
%%%%%%%%%%%%%%%%%%%%%%%%%%%%%%%%
%%%%%%%%%%%%%%%%%%%%%%%%%%%%%%%%
%%%%%%%%%%%%%%%%%%%%%%%%%%%%%%%%
%%%%%%%%%%%%%%%%%%%%%%%%%%%%%%%%
%%%%%%%%%%%%%%%%%%%%%%%%%%%%%%%%
%%%%%%%%%%%%%%%%%%%%%%%%%%%%%%%%
%%%%%%%%%%%%%%%%%%%%%%%%%%%%%%%%
%%%%%%%%%%%%%%%%%%%%%%%%%%%%%%%%
%%%%%%%%%%%%%%%%%%%%%%%%%%%%%%%%
\subsection{Linear diffusive modes}
%%%%%%%%%%%%%%%%%%%%%%%%%%%%%%%%
%%%%%%%%%%%%%%%%%%%%%%%%%%%%%%%%
%%%%%%%%%%%%%%%%%%%%%%%%%%%%%%%%
%%%%%%%%%%%%%%%%%%%%%%%%%%%%%%%%
%%%%%%%%%%%%%%%%%%%%%%%%%%%%%%%%
%%%%%%%%%%%%%%%%%%%%%%%%%%%%%%%%
%%%%%%%%%%%%%%%%%%%%%%%%%%%%%%%%
%%%%%%%%%%%%%%%%%%%%%%%%%%%%%%%%
%%%%%%%%%%%%%%%%%%%%%%%%%%%%%%%%
In the linear case for $\lambda = 0$ the Hamiltonian $H$ 
is 
harmonic in the fields $u$ and $\varphi$. The coupled field equations
are linear,

%{\em l1, l2}
%
\begin{eqnarray}
\frac{\partial u}{\partial t}&&
=-i\nu\nabla^2\varphi 
\label{l1}
\\
\frac{\partial\varphi}{\partial t} &&
=+i\nu\nabla^2 u ~ ,  
\label{l2}
\end{eqnarray}
and describe the weak noise limit of the EW equation.
Expanding about the stationary state
$(u_0,\varphi_0) = (0,0)$ the equations readily admit the solutions

%{\em uexp, fiexp}
%
\begin{eqnarray}
u(xt)&&=\sum_k[
u_k^{(+)}e^{-i\omega_k^0 t+ikx}+
u_k^{(-)}e^{+i\omega_k^0 t-ikx}] 
\label{uexp}
\\
\varphi(xt)&&=i \sum_k[
u_k^{(+)}e^{-i\omega_k^0 t+ikx}-
u_k^{(-)}e^{+i\omega_k^0 t-ikx}] 
\label{fiexp}
\end{eqnarray}
with the quadratic diffusive dispersion law

%{\em disp3}
%
\begin{equation}
\omega_k^0 = -i\nu k^2 ~ .
\label{disp3}
\end{equation}
Since $u$ is real we have $(u_k^{(\pm)})^\ast = u_{-k}^{(\pm)}$
implying that $\varphi$ is purely imaginary as discussed above. We also
note that the solution (\ref{uexp}), unlike the solution of the noiseless
EW equation, includes both 
growing and
decaying solutions. As discussed earlier this feature is 
consistent with the
time reversal invariance
in the stationary regime.
%%%%%%%%%%%%%%%%%%%%%%%%%%%%%%%%
%%%%%%%%%%%%%%%%%%%%%%%%%%%%%%%%
%%%%%%%%%%%%%%%%%%%%%%%%%%%%%%%%
%%%%%%%%%%%%%%%%%%%%%%%%%%%%%%%%
%%%%%%%%%%%%%%%%%%%%%%%%%%%%%%%%
%%%%%%%%%%%%%%%%%%%%%%%%%%%%%%%%
%%%%%%%%%%%%%%%%%%%%%%%%%%%%%%%%
%%%%%%%%%%%%%%%%%%%%%%%%%%%%%%%%
%%%%%%%%%%%%%%%%%%%%%%%%%%%%%%%%
\subsection{Nonlinear soliton modes}
%%%%%%%%%%%%%%%%%%%%%%%%%%%%%%%%
%%%%%%%%%%%%%%%%%%%%%%%%%%%%%%%%
%%%%%%%%%%%%%%%%%%%%%%%%%%%%%%%%
%%%%%%%%%%%%%%%%%%%%%%%%%%%%%%%%
%%%%%%%%%%%%%%%%%%%%%%%%%%%%%%%%
%%%%%%%%%%%%%%%%%%%%%%%%%%%%%%%%
%%%%%%%%%%%%%%%%%%%%%%%%%%%%%%%%
%%%%%%%%%%%%%%%%%%%%%%%%%%%%%%%%
%%%%%%%%%%%%%%%%%%%%%%%%%%%%%%%%
In order to treat the nonlinear aspects of the field equations 
we employ the same method as in the analysis of the noiseless
Burgers equation  and look for
static solutions. Using the Galilean invariance 
propagating solutions are then easily generated by a transformation to a moving
frame accompanied by a shift of $u$.
The static case, 
$\partial u/\partial t = \partial\varphi/\partial t = 0$, corresponds
to the action $S = -\int dxdt{\cal H}(u,\varphi)$ and the solutions are 
given by the stationary points of the Hamiltonian $H$.
Multiplying the
static field equations by $\nabla\varphi$ and $\nabla u$, respectively, we
obtain $\nabla\varphi\nabla^2\varphi + \nabla u\nabla^2 u =0$, or by quadrature,
imposing the boundary conditions of vanishing slope,
$\nabla u = \nabla\varphi = 0$ for $x\rightarrow\pm L/2$, the 
slope condition

%{\em slcon}
%
\begin{equation}
(\nabla\varphi)^2 + (\nabla u)^2 = 0 ~ .
\label{slcon}
\end{equation}
Solving the slope condition we obtain

%{\em tsol}
%
\begin{equation}
\nabla\varphi=i\mu\nabla u
\label{tsol}
\end{equation}
parametrized by the parity index $\mu=\pm 1$, which inserted in the static field
equations yields

%{\em sfe}
%
\begin{equation}
\mu\nu\nabla^2 u + \lambda u\nabla u =0 ~ .
\label{sfe}
\end{equation}
This equation has the same form as the static limit of the noiseless 
Burgers equation (\ref{nlbur}) with damping $\mu\nu$,
and we obtain the 
static solution (\ref{ssol}) with $\nu$ replaced by
$\mu\nu$, i.e., $\pm\nu$,

%{\em stsol}
%
\begin{equation}
u(x) =u_+
\tanh{\left[\frac{\mu\lambda u_+}{2\nu}(x-x_0)\right]} ~ ,
\label{stsol}
\end{equation}
The solution (\ref{stsol}) has the form of a  static, 
localized, symmetric
soliton or kink with amplitude $2|u_+|$, center of mass $x_0$,
width $2\nu/(\lambda|u_+|)$, approaching
$\pm\mu|u_+|$ for $x\rightarrow\pm L/2$. In the limit of vanishing
damping, $\nu\rightarrow 0$, the soliton becomes a sharp discontinuity
or shock in the slope field.
The static soliton connects two degenerate stationary states with
slopes $\pm|u_+|$. However, unlike the sine-Gordon soliton 
\cite{scott,rajaraman}
which is characterized by a topological quantum number or the
$\varphi^4$ soliton \cite{scott,rajaraman} which connects the two degenerate
ground states defined by the double-well potential, the Burgers soliton
has an arbitrary amplitude $|u_+|$ corresponding to the infinitely
degenerate stationary states.
Furthermore, we note 
the interesting fact that unlike the case of the noiseless
Burgers equation, 
where we only have a single {\em right hand} soliton mode,
corresponding to $\mu = +1$, the broken reflection or
parity symmetry is restored in the noisy case.
The noise drives the interface into a stationary state and in the process
excites both {\em right} and {\em left hand} solitons. This mechanism
in the nonlinear case is equivalent to the excitation of both growing and
decaying diffusive modes in the EW case.

The static soliton is a special configuration connecting stationary
states with opposite slopes. However, since the underlying field
equations are invariant under the Galilean symmetry group 
(\ref{gal1}-\ref{gal2}) it is an easy task to construct 
a 
propagating soliton
solutions by means of a Galilean transformation.
Similar to the discussion of the noiseless Burgers equation we obtain,
introducing the boundary values $u_\pm$ for $x\rightarrow L/2$, the 
soliton condition (\ref{solcon}), i.e., 

%{\em solcon2}
%
\begin{equation}
u_+ + u_- =-\frac{2v}{\lambda} ~ .
\label{solcon2}
\end{equation}
and the
moving soliton solution 

%{\em mvsol}
%
\begin{equation}
u(xt) = \frac{u_++u_-}{2} + \frac{u_+-u_-}{2}
\tanh{\left[\frac{\lambda}{4\nu}|u_+-u_-|(x-vt-x_0)\right]} ~ ,
\label{mvsol}
\end{equation}
connecting stationary states with slopes $u_+$ and $u_-$; note that we
have absorbed the parity index $\mu$ in the sign of $u_+ - u_-$. 
The soliton has center of mass $x_0$, propagates with velocity
$v = -\lambda(u_++u_-)/2$, has the width $4\nu/(\lambda|u_+-u_-)$,
and amplitude $|u_+-u_-|$. In the limit of vanishing damping 
$\nu\rightarrow 0$,
the inviscid limit, or large drive $\lambda$, the soliton becomes a shock
wave, i.e., a discontinuity in the slope field; for finite damping the 
shock wave front is smoothed by dissipation. For vanishing amplitude,
$u_+-u_-\rightarrow 0$, the soliton merges continuously into the 
stationary
state $u_+ = u_-$.

Finally, integrating the slope condition (\ref{slcon}), which by inspection 
also
holds for the propagating soliton, and using the invariance property
(\ref{fishift}) in order
to set the integration constant equal to zero, we obtain

%{\em fiurel}
%
\begin{equation}
\varphi = i\mu u ~ ,
\label{fiurel}
\end{equation}
yielding the static and propagating soliton solutions for the associated
noise field and in accordance with the saddle point regions in Fig. 7. 
In Fig. 8 we have depicted the static solitons and the
associated smoothed cusps in the height field.
%%%%%%%%%%%%%%%%%%%%%%%%%%%%%%%
%%%%%%%%%%%%%%%%%%%%%%%%%%%%%%%
%%%%%%%%%%%%%%%%%%%%%%%%%%%%%%%
%%%%%%%%%%%%%%%%%%%%%%%%%%%%%%%
%%%%%%%%%%%%%%%%%%%%%%%%%%%%%%%
%%%%%%%%%%%%%%%%%%%%%%%%%%%%%%%
\subsection{Multi-soliton solutions - Boundary conditions}
%%%%%%%%%%%%%%%%%%%%%%%%%%%%%%%
%%%%%%%%%%%%%%%%%%%%%%%%%%%%%%%
%%%%%%%%%%%%%%%%%%%%%%%%%%%%%%%
%%%%%%%%%%%%%%%%%%%%%%%%%%%%%%%
%%%%%%%%%%%%%%%%%%%%%%%%%%%%%%%
%%%%%%%%%%%%%%%%%%%%%%%%%%%%%%%
In addition to defining the path integral in a finite LT box,
we must also specify appropriate
boundary conditions for the slope and noise fields in accordance with
the physical situation. For a growing interface it is convenient to assume
a horizontal interface at the boundaries equivalent to a vanishing
slope field. This corresponds to a vanishing deterministic component
in the current in Eq. (\ref{cur2}) at the boundaries and implies that
the interface is only driven by the noise.
However, since the single soliton solution discussed above connect stationary
states with different slopes corresponding to a non-vanishing current,
we must pair at least two solitons of opposite parity in order to
satisfy the boundary conditions. By inspection of the field equations
(\ref{eq1}-\ref{eq2}) we note, however, that a non-overlapping
two-soliton configuration $u^{(1)}+u^{(2)}$ connected by a segment
of constant slope is an approximate solution to the field equations
and therefore corresponds to an extremum of the action in the path
integral. The correction term is given by
$\lambda(u^{(1)}\nabla u^{(2)} + u^{(2)}\nabla u^{(1)})$
whose contribution to the action we can ignore for non-overlapping
well-separated solitons. Furthermore, the argument can be generalized
to a multi-soliton configuration connected by segments of constant slope.
It is a well-known feature of path integral instanton or solitons configurations
that one in order to obtain the correct asymptotic behavior must sum over
a gas of non-overlapping instantons or solitons \cite{das,rajaraman}.
The situation is the same
in the present somewhat more complicated context. In order to satisfy the
boundary conditions of vanishing slope and to collect all the leading
contributions in the asymptotic weak noise limit the structure of the
path integral implies the formation of a
dilute gas of non-overlapping solitons.
In Fig. 9 we have shown the case of two non-overlapping soliton solutions.
%%%%%%%%%%%%%%%%%%%%%%%%%%%%%%%
%%%%%%%%%%%%%%%%%%%%%%%%%%%%%%%
%%%%%%%%%%%%%%%%%%%%%%%%%%%%%%%
%%%%%%%%%%%%%%%%%%%%%%%%%%%%%%%
\section{Dynamics of solitons - Principle of least action}
%%%%%%%%%%%%%%%%%%%%%%%%%%%%%%%
%%%%%%%%%%%%%%%%%%%%%%%%%%%%%%%
%%%%%%%%%%%%%%%%%%%%%%%%%%%%%%%
%%%%%%%%%%%%%%%%%%%%%%%%%%%%%%%
%%%%%%%%%%%%%%%%%%%%%%%%%%%%%%%
%%%%%%%%%%%%%%%%%%%%%%%%%%%%%%%
It is a fundamental aspect of the canonical form of the path integral
for the noisy Burgers equation that it supports 
{\em a principle of least action}
\cite{landau1}. 
Therefore, unlike the noiseless Burgers equation where
there is no underlying canonical structure,
the asymptotic weak noise soliton and diffusive mode  solutions
are derived from a variational principle. Furthermore, we can associate an 
effective action,
energy, and momentum with a particular phase space configuration
or growth pattern.

The energy density $\epsilon$ is generally given by Eq. (\ref{ham}). 
Inserting the
slope condition (\ref{slcon}) valid for the soliton solutions, the harmonic
part of $\epsilon$ cancels and the soliton solutions exclusively contribute
to the growth term, i.e., $\epsilon=(\lambda/2)u^2\nabla\varphi$, or in
terms of the momentum density (\ref{md}), $\epsilon = (\lambda/2)ug$. Inserting the
soliton constraint (\ref{tsol})  the energy density also takes the form

%{\em ed}
%
\begin{equation}
\epsilon = (\lambda/2)i\mu u^2\nabla u ~.
\label{ed}
\end{equation}
We note that the energy density is localized to the position 
of the soliton where $u$ varies most rapidly. For the  soliton energy
we thus obtain by quadrature in terms of the boundary values $u_\pm$,

%{\em e}
%
\begin{equation}
E = i\mu\frac{\lambda}{6}[u_+^3-u_-^3] ~ .
\label{e}
\end{equation}
In a similar manner the soliton momentum (\ref{mom}) is given by

%{\em p}
%
\begin{equation}
P = i\mu\frac{1}{2}[u_+^2 -u_-^2]
\label{p}
\end{equation}
and finally from Eq. (\ref{ac}), using 
$\partial\varphi/\partial t = -v\nabla\varphi$
for the boosted static soliton, the soliton action

%{\em a}
%
\begin{equation}
S = -T[Pv +E] ~ ,
\label{a}
\end{equation}
which also follows from
the Galilean invariance of $S$ \cite{landau1}.

The purely imaginary character of $E$, $P$, and $S$ is a
feature of our choice of convention in establishing the canonical 
path integral in Eqs. (\ref{gen3}-\ref{ham}).
In order to exploit the formalism of analytical mechanics and the 
structure of the phase 
space Feynman path integral we have chosen the noise field $\varphi$ in such
a manner that $u$ and $\varphi$ appear as canonically conjugate 
variables satisfying
the usual Poisson bracket  (\ref{pa}). As discussed earlier this 
implies
that $\varphi$  has to be purely imaginary in the case of the
weak noise saddle point solutions.
The
complex character of $E$ and $P$ is consistent with the relaxational
and propagating aspects of the soliton modes as also observed in the 
Fokker-Planck description in L. The main properties of 
$E$ and $P$ in the
present dynamical context are that they serve as generators of translations in
time and space, respectively \cite{landau1}.
The nonlinear energy-momentum relationship
is  characteristic of nonlinear soliton solutions
\cite{fogedby6} and is different from the simple $E\text{-}P$ relationship 
encountered in the Lorentz invariant
$\varphi^4$ or sine-Gordon equation \cite{scott}.
We also note that the damping constant $\nu$ does not enter in the 
expressions
for $E$ and $P$ which only depend on the boundary values 
$u_\pm$
and the drive $\lambda$. In the weak noise limit the dynamics of soliton
solutions is thus entirely decoupled from the dynamics of the linear 
diffusive modes.

Let us specifically consider a soliton configuration satisfying the 
boundary condition
of left vanishing slope, i.e.,  $u_- = 0$ for $x=-L/2$. The soliton condition 
(\ref{solcon2})
then implies the right boundary value $u_+=-2v/\lambda$,
relating $u_+$ to the propagation velocity $v$. The soliton with positive
parity, $u_+>0$, propagates left with negative velocity; whereas the
soliton with opposite parity, $u_+<0$, propagates in the forward direction.
For  $E$ and $P$ we infer 
for both parities

%{\em e1, p1, s1}
%
\begin{eqnarray}
E = && \frac{4}{3}i\frac{|v|^3}{\lambda^2}
\label{e1}
\\
P = && -2i\frac{v|v|}{\lambda^2}
\label{p1}
\\
S = && \frac{2}{3}iT\frac{|v|^3}{\lambda^2} ~ .
\label{s1}
\end{eqnarray}
The velocity $v=-\lambda u_+/2$ characterizes the kinematics of the soliton
and is related to the amplitude whereas $E$ and $P$ determine the 
transformation properties. Eliminating the velocity we obtain the
soliton dispersion law

%{\em emdisp, fase}
%
\begin{equation}
E_P = i\lambda\frac{\sqrt{2}}{3}|P|^\frac{3}{2} ~ ,
\label{emdisp}
\end{equation}
where $\text{sign}(\text{Im}P)=-\text{sign}v$.
We note that the nonlinear 
localized soliton excitation has a {\em qualitatively} different 
dispersion law from the linear extended diffusive mode dispersion law
$\omega=-i\nu k^2$.
They are both gapless modes but the exponents are different.
The consequences of this aspect on the spectrum and scaling properties
will be investigated later when we consider the fluctuations in more
detail, but we already note here that the change in the exponent which
can be identified with the dynamic exponent $z$  is related to the
different universality classes for the EW and Burgers cases.

Since the energy and momentum densities are localized at the soliton
positions it follows that they are additive quantities for a
multi-soliton configuration and the general expressions
(\ref{e}),(\ref{p}), and (\ref{a}) allow us to evaluate the total energy, momentum, and
action for an arbitrary configuration constructed from well-separated
non-overlapping solitons.

%%%%%%%%%%%%%%%%%%%%%
%%%%%%%%%%%%%%%%%%%%%
%%%%%%%%%%%%%%%%%%%%%
%%%%%%%%%%%%%%%%%%%%%
%%%%%%%%%%%%%%%%%%%%%
%%%%%%%%%%%%%%%%%%%%%
\section{A growing interface as a dilute soliton gas}
%%%%%%%%%%%%%%%%%%%%%
%%%%%%%%%%%%%%%%%%%%%
%%%%%%%%%%%%%%%%%%%%%
%%%%%%%%%%%%%%%%%%%%%
%%%%%%%%%%%%%%%%%%%%%
%%%%%%%%%%%%%%%%%%%%%
%%%%%%%%%%%%%%%%%%%%%
%%%%%%%%%%%%%%%%%%%%%
%%%%%%%%%%%%%%%%%%%%%
We are now in position to present a coherent picture of the morphology
of a statistically driven growing interface. In the weak noise limit
$\Delta\rightarrow 0$ {\em the principle of least action}  which operates
in the present context implies that the stationary points in $(u,\varphi)$
phase space correspond to solitons and multi-soliton configurations
connected by segments of constant slope. In addition there will be superposed
diffusive modes. In the nonlinear case the soliton configurations
determine the dominant features of the growth morphology and will
be considered here. The superposed diffusive modes will be discussed
in the next section.

The weight of a particular soliton configuration in the path integral
is given by the action which is an additive quantity for a dilute
gas of solitons. The soliton configurations are assumed to be excited
with respect to a stationary state of vanishing slope, i.e., a
horizontal interface, and are furthermore determined by imposing
periodic boundary  conditions at $x=\pm L/2$; we remark that fixed
boundary conditions are inconsistent with a soliton configuration
moving across the system.
We also note that unlike the transient properties of the noiseless
Burgers equation which are described by a gas of 
{\em right hand} solitons
connected by ramps, corresponding to a transient height profile 
composed of
smoothed cusps connected by convex parabolic segments as shown in Fig. 5 ,
the stationary
state of the noisy Burgers equation is characterized by a gas of both 
{\em right} and {\em left hand} solitons connected by pieces of constant slope. 
The noise thus radically changes the growth morphology of the Burgers 
equation. The noise stochastically modifies the transient regime by
exciting solitons of both parities which thus describe the morphology
of the slope field in the stationary nonequilibrium state.
The situation is similar to the case of the noise-driven
damped sine-Gordon equation \cite{buttiker,krug4} where the noise also
excites nonlinear soliton modes.

We now proceed to discuss the morphology of a growing interface in terms
of solitons in the slope field $u$. 
The basic ``building block'' is the
{\em static soliton} configuration given by Eq. (\ref{stsol}). 
From Eqs. (\ref{e}-\ref{a})
it follows that this mode has vanishing momentum $P=0$, energy
$E=i(\lambda/3)|u_+|^3$, and action $S=-i(\lambda/3)T|u_+|^3$, independent of
its parity. By integration the height profile, $h = \int udx$, is given by

%{\em hfield}
%
\begin{equation}
h(x) = \frac{2\nu}{\lambda}\frac{u_+}{|u_+|}
\log{\cosh{\left[\frac{\lambda u_+}{2\nu}(x-x_0)\right]}} ~ ,
\label{hfield}
\end{equation}
corresponding to a downward and an upward pointing cusp smoothed by the
damping constant $\nu$. In the limit of vanishing $\nu$ the cusps become
sharp. We also note that the static soliton does not satisfy the boundary
conditions of vanishing slope. In Fig. 8 we have depicted the slope and
height field in the two cases.

Boosting the static soliton in Eq. (\ref{stsol}) we obtain a 
{\em single moving soliton}
with velocity given by the soliton condition (\ref{solcon2}) and the profile
(\ref{mvsol}). In the particular case of a soliton satisfying the left boundary
condition $u_-=0$, we obtain from Eq. (\ref{mvsol}) the slope field

%{\em ufield}
%
\begin{equation}
u(xt)=\frac{u_+}{2}
\left[1+\tanh{\left[\frac{\lambda|u_+|}{4\nu}(x-vt-x_0)\right]}\right]
\label{ufield}
\end{equation}
and by integration the height profile

%{\em hfield2}
%
\begin{equation}
h(xt) = \frac{u_+}{2}x+\frac{2\nu}{\lambda}\frac{u_+}{|u_+|}
\log{\cosh{\left[\frac{\lambda|u_+|}{4\nu}(x-vt-x_0)\right]}}
\label{hfield2}
\end{equation}
with propagation velocity

%{\em propv}
%
\begin{equation}
v = -\frac{\lambda u_+}{2} ~ .
\label{propv}
\end{equation}
The energy, momentum, and action are given by Eqs. (\ref{e1}-\ref{s1}).
This  mode corresponds to the bottom part of an ascending step or top part
of a descending step in the height field propagating to the left or right,
depending on the sign of $u_+$. The configurations are shown in Fig. 10.

In order to describe a {\em moving step} in the height profile
we 
pair two well-separated non-overlapping solitons with equal 
amplitude and opposite
parity. The soliton condition (\ref{solcon2}) then implies that they move in
the same direction with the same velocity. In this case the slope and height
fields have the form

%{\em uu,hh,vv}

%
\begin{eqnarray}
u(xt) =&& \frac{u_+}{2}
\left[
\tanh{\frac{\lambda|u_+|}{4\nu}(x-vt-x_1)}
-
\tanh{\frac{\lambda|u_+|}{4\nu}(x-vt-x_2)}
\right]
\label{uu}
\\
h(xt) =&& \frac{2\nu}{\lambda}
\frac{u_+}{|u_+|}
\log
{
\left[
\frac
{\cosh{(\lambda|u_+|/4\nu)(x-vt-x_1)}}
{\cosh{(\lambda|u_+|/4\nu)(x-vt-x_1)}}
\right]
}
\label{hh}
\\
v=&&-\frac{\lambda u_+}{2} ~ .
\label{vv}
\end{eqnarray}
We have here assumed $x_1\ll x_2$ for the center of mass coordinates.
This configuration corresponds to two co-moving solitons moving with velocity
$v=-\lambda u_+/2$ and is equivalent to a moving step in the height profile.
The height of the step $\Delta h$ is given by 
$\Delta h = u_+(x_2-x_1)$. Imposing periodic boundary conditions for the 
slope field corresponding to a closed ring of length $L$, this two-soliton 
mode corresponds to a step in $h$ moving along the closed ring. 
At each revolution
the height field thus increases by $\Delta h$ and we have a simple 
growth situation.
For well-separated solitons the energy, momentum, and 
action are additive and we obtain from Eqs. (\ref{e}-\ref{a})

%{\em eee,ppp,sss}
%
\begin{eqnarray}
E_{step}=&&\frac{8}{3}i\frac{|v|^3}{\lambda^2}
\label{eee}
\\
P_{step}=&&-4i\frac{v|v|}{\lambda^2}
\label{ppp}
\\
S_{step}=&&\frac{4}{3}iT\frac{|v|^3}{\lambda^2} ~ .
\label{sss}
\end{eqnarray}
In Fig. 11 we have shown the configurations.

In a similar way we can construct a more faceted height profile in terms
of a gas of appropriately paired solitons in the slope field $u$ with the
only requirement that i) the solitons are well-separated so that they
constitute saddle point solutions and ii) they satisfy periodic
boundary conditions.
For example a {\em growing tip}
or the {\em filling in} of an indentation is described by the 
three-soliton configurations. A {\em growing plateau}
formed by two step corresponds to a four-soliton configuration.
We also notice that the two-soliton configurations corresponding to a 
moving step can be ``renormalized'' by the excitation of further two-soliton
configurations corresponding to curvature of
the step. In Fig. 12 we have depicted the above special
configurations. In Fig. 13 we have shown a general profile.
%%%%%%%%%%%%%%%%%%%%%
%%%%%%%%%%%%%%%%%%%%%
%%%%%%%%%%%%%%%%%%%%%
%%%%%%%%%%%%%%%%%%%%%
%%%%%%%%%%%%%%%%%%%%%
%%%%%%%%%%%%%%%%%%%%%
%%%%%%%%%%%%%%%%%%%%%
\section{``Quantum description'' of a growing interface -- fluctuations}
%%%%%%%%%%%%%%%%%%%%%
%%%%%%%%%%%%%%%%%%%%%
%%%%%%%%%%%%%%%%%%%%%
%%%%%%%%%%%%%%%%%%%%%
%%%%%%%%%%%%%%%%%%%%%
%%%%%%%%%%%%%%%%%%%%%
%%%%%%%%%%%%%%%%%%%%%
In the previous section we demonstrated that the dominant morphology 
of a growing
interface governed by the noisy Burgers equation in the weak noise limit
can be described in terms of a dilute gas of propagating solitons. 
In the
path integral
the soliton contributions correspond to the stationary saddle points
in the $(u,\varphi)$ phase space determined by the principle of least
action. By inspection of the one dimensional integral 
(\ref{int}) 
and the saddle point result (\ref{int2}), it is clear that the 
soliton solution 
corresponds to
the stationary point $u_0$ and the associated soliton action to $S(u_0)$. 
Consequently, we have
not included the Gaussian fluctuations about the stationary point,
yielding the multiplicative factor in Eq. (\ref{int2}) of order 
$\Delta^{1/2}$ and
depending on the second order derivative $S''(u_0)$ evaluated at the 
stationary point, but only taken into account the exponential
contribution determined by the action.
In the context of the path integral the Gaussian fluctuations about the
stationary soliton  correspond to the linear diffusive mode 
spectrum in the presence of the soliton configurations and remains to
be discussed.

In order to proceed in the analysis of the path integral representation
of the noisy Burgers equation we shall 
take
a heuristic point of view and extract some information and physical
insight by making use of 
the Feynman
path integral structure of $Z$ in Eqs. (\ref{gen3}-\ref{ham})),
deferring an analysis of the path integral {\em per se}
to another context. The idea
is to in a certain sense ``deconstruct''
the path integral and determine the form of the underlying
``quantum field theory'' leading to $Z$ by the usual Feynman method 
\cite{feynman,das,rajaraman}.
Since the slope field $u$ and the noise field $\varphi$ in the path integral
form a canonically conjugate pair with Poisson bracket (\ref{pa}), where 
$u$ plays the role of a
canonical ``momentum'' and $\varphi$  a canonically conjugate 
``coordinate'', 
the first step, is to 
introduce the ``quantum fields'' 
$\hat{u}$ and $\hat{\varphi}$ 
satisfying the canonical commutator

%{\em cancom}
%
\begin{equation}
[\hat{u}(x),\hat{\varphi}(x')]=-i\frac{\Delta}{\nu}\delta(x-x') ~ .
\label{cancom}
\end{equation}
Here the ratio of the noise to the damping, $\Delta/\nu$, plays the role
of an effective Planck constant just as in the path integral. 
We thus have an effective ``correspondence 
principle'' operating relating the ``classical'' Poisson bracket $\{A,B\}$ 
to the ``quantum''
commutator $[\hat{A},\hat{B}]$, according to the prescription
$[\hat{A},\hat{B}]=-i(\Delta/\nu)\{A,B\}$. In a similar way the effective
``quantum Hamiltonian'' $\hat{H}$ is inferred from (\ref{ha}) 
\cite{footnote5},

%{\em qh}
%
\begin{equation}
\hat{H} = \int\left[-i\frac{\nu}{2}[(\nabla\hat{u})^2+(\nabla\hat{\varphi})^2]
+\frac{\lambda}{2}\hat{u}^2\nabla\hat{\varphi}\right]dx ~ .
\label{qh}
\end{equation}

Whereas the fields $\hat{u}$ and $\hat{\varphi}$ by construction are
Hermitian the Hamiltonian $\hat{H}$ is in general a non-Hermitian operator. 
Expressing $\hat{H}$ in the form 
$\hat{H}_0+\hat{H}_1$ it is composed of an anti-Hermitian harmonic component
$\hat{H}_0$, governing the dynamics of the linear diffusive modes and a 
nonlinear Hermitian component
$\hat{H}_1$, describing the growth characterized by $\lambda$.
In the Heisenberg picture  $\hat{H}$ is the generator of time translations 
and we obtain the
usual Heisenberg equations of motion \cite{footnote6} 

%{\em heis1, heis2}
%
\begin{eqnarray}
\frac{\partial\hat{u}}{\partial t} = && i\frac{\nu}{\Delta}[\hat{H},\hat{u}]
\label{heis1}
\\
\frac{\partial\hat{\varphi}}{\partial t} = 
&& i\frac{\nu}{\Delta}[\hat{H},\hat{\varphi}] ~ ,
\label{heis2}
\end{eqnarray}
yielding ``quantum field equations'' of the same form as the ``classical''
field equations (\ref{eq1}-\ref{eq2}) \cite{footnote7},

%{\em clf1, clf2}
%
\begin{eqnarray}
\frac{\partial\hat{u}}{\partial t} = && -i\nu\nabla^2\hat{\varphi}
+ \lambda\hat{u}\nabla\hat{u}
\label{clf1}
\\
\frac{\partial\hat{\varphi}}{\partial t} = && +i\nu\nabla^2\hat{u}
+ \lambda\hat{u}\nabla\hat{\varphi} ~ .
\label{clf2}
\end{eqnarray}
In a similar way, the momentum operator $\hat{P}$, the generator of
translation, is given by

%{\em momen}
%
\begin{equation}
\hat{P} = \int dx\hat{u}\nabla\hat{\varphi} ~ ,
\label{momen}
\end{equation}
giving rise to the commutator relations

%{\em comm1, comm2}
%
\begin{eqnarray}
\nabla\hat{u} =&& i\frac{\nu}{\Delta}[\hat{P},\hat{u}]
\label{comm1}
\\
\nabla\hat{\varphi} =&& i\frac{\nu}{\Delta}[\hat{P},\hat{\varphi]} ~ .
\label{comm2}
\end{eqnarray}
Finally, the symmetry algebra in section V also holds in the
``quantum case'' by simply  replacing the Poisson
brackets by commutators according to the above ``correspondence principle''.

The ``quantum field equations'' (\ref{clf1}-\ref{clf2})) together 
with the appropriate states
of the Hamiltonian (\ref{qh}) are completely equivalent to the path integral
and thus provide an alternative  description of the noisy Burgers equation. 
The noise-induced 
fluctuations in the slope field, represented by the different 
configurations or paths in the path integral weighted by the 
``classical''
action $S$,  are replaced by  ``quantum fluctuations''
in the underlying ``quantum field theory'', resulting from the operator 
structure and the associated commutator algebra.
We also note that the ``quantum description'' presented here is precisely the
same as the one obtained in L  based on the 
mapping of a
solid-on-solid model to a continuum spin chain model in the quasi-classical
limit.
%%%%%%%%%%%%%%%%%%%%%%%%%%%%%%%%%%%%%%%%%%%%%
%%%%%%%%%%%%%%%%%%%%%%%%%%%%%%%%%%%%%%%%%%%%%
%%%%%%%%%%%%%%%%%%%%%%%%%%%%%%%%%%%%%%%%%%%%%
%%%%%%%%%%%%%%%%%%%%%%%%%%%%%%%%%%%%%%%%%%%%%
%%%%%%%%%%%%%%%%%%%%%%%%%%%%%%%%%%%%%%%%%%%%%
%%%%%%%%%%%%%%%%%%%%%%%%%%%%%%%%%%%%%%%%%%%%%
%%%%%%%%%%%%%%%%%%%%%%%%%%%%%%%%%%%%%%%%%%%%%
%%%%%%%%%%%%%%%%%%%%%%%%%%%%%%%%%%%%%%%%%%%%%
\subsection{The Edwards-Wilkinson equation}
%%%%%%%%%%%%%%%%%%%%%%%%%%%%%%%%%%%%%%%%%%%%%
%%%%%%%%%%%%%%%%%%%%%%%%%%%%%%%%%%%%%%%%%%%%%
%%%%%%%%%%%%%%%%%%%%%%%%%%%%%%%%%%%%%%%%%%%%%
%%%%%%%%%%%%%%%%%%%%%%%%%%%%%%%%%%%%%%%%%%%%%
%%%%%%%%%%%%%%%%%%%%%%%%%%%%%%%%%%%%%%%%%%%%%
%%%%%%%%%%%%%%%%%%%%%%%%%%%%%%%%%%%%%%%%%%%%%
%%%%%%%%%%%%%%%%%%%%%%%%%%%%%%%%%%%%%%%%%%%%%
%%%%%%%%%%%%%%%%%%%%%%%%%%%%%%%%%%%%%%%%%%%%%
In order to demonstrate how the ``quantum scheme'' works it is
instructive to evaluate the slope correlation function 
$\langle uu\rangle(k\omega)$ in Eq. (\ref{lor}) for the Edwards-Wilkinson
equation. The dynamics of the EW case is governed by the unperturbed part of
$\hat{H}$ in Eq. (\ref{qh})

%{\em ewham}
%
\begin{equation}
\hat{H}_0 = \int\left[-i\frac{\nu}{2}[(\nabla\hat{u})^2
+
(\nabla\hat{\varphi})^2]\right]dx ~ .
\label{ewham}
\end{equation}
Introducing the usual ``second quantization'' scheme \cite{mahan}
in terms of Bose
annihilation and creation operators $a_k$ and $a^\dagger_k$ satisfying
the commutator algebra $[a_k,a^\dagger_k] = \delta_{kk'}$, we have for the
slope and noise fields
%

%{\em opu,opfi}
\begin{eqnarray}
\hat{u} =&& -i\sqrt{\frac{\Delta}{2\nu L}}
\sum_k[e^{ikx}a_k - e^{-ikx}a^\dagger_k]
\label{opu}
\\
\hat{\varphi} =&& \sqrt{\frac{\Delta}{2\nu L}}
\sum_k[e^{ikx}a_k + e^{-ikx}a^\dagger_k] ~ ,
\label{opfi}
\end{eqnarray}
yielding the unperturbed Hamiltonian and associated diffusive dispersion
law
%

%{\em hup}
\begin{eqnarray}
\hat{H}_0 =&& \sum_k\frac{\Delta}{\nu}\omega_k^0a^\dagger_ka_k
\label{hup}
\\
\omega_k^0 = && -i\nu k^2 ~ .
\label{d1}
\end{eqnarray}
Noting that the particle vacuum state $|0\rangle$ 
corresponds to a  stationary
state with average vanishing slope $\langle 0|\hat{u}|0\rangle$, i.e.,
a horizontal interface,
it is an easy task to evaluate the slope
correlation function. Since the path integral defines a Bose time-ordering
\cite{das} and using the time evolution operator, we have the identification
%

%{\em g1}
\begin{equation}
\langle u(x,t)u(0,0)\rangle =
\langle 0|T\hat{u}(x,t)\hat{u}(0,0)|0\rangle =
\langle 0|\hat{u}(x)e^{-i\hat{H}_0|t|/(\Delta/\nu)}\hat{u}(0)|0\rangle ~ .
\label{g1}
\end{equation}
Using that $a_k$ 
evolves in time according to

%{\em cop}
%
\begin{equation}
a_k(t) = a_k(0)\exp{\left(-i\frac{\Delta}{\nu}\omega_k^0t\right)}
\label{cop}
\end{equation}
we then obtain in Fourier space,

%{\em co3}
%
\begin{equation}
\langle u(k,\omega)u(-k,-\omega)\rangle = i\frac{\Delta}{2\nu}
\left[
\frac
{
\langle 0|a_ka_k^\dagger|0\rangle
}
{
\omega_k^0-\omega
}
+
\frac
{
\langle 0|a_ka_k^\dagger|0\rangle
}
{
\omega_k^0+\omega
}
\right]
\label{co3}
\end{equation}
or in reduced form in complete agreement with Eq. (\ref{lor}), 

%{\em co2}
%
\begin{equation}
\langle u(k,\omega)u(-k,-\omega)\rangle =
\frac{\Delta k^2}{\omega^2 - (\omega_k^0)^2} ~ .
\label{co2}
\end{equation}

This simple calculation  demonstrates
how the ``quantum fluctuations'' as expressed by the commutator
algebra  and the effective Planck constant $\Delta/\nu$
combine to produce the factor $\Delta$ in
the correlation function which ``classically''
in terms of the EW Langevin equation originates from averaging
over the noise $\nabla\eta$.
%%%%%%%%%%%%%%%%%%%%%%%%%%%%%%%
%%%%%%%%%%%%%%%%%%%%%%%%%%%%%%%
%%%%%%%%%%%%%%%%%%%%%%%%%%%%%%%
%%%%%%%%%%%%%%%%%%%%%%%%%%%%%%%
%%%%%%%%%%%%%%%%%%%%%%%%%%%%%%%
%%%%%%%%%%%%%%%%%%%%%%%%%%%%%%%
%%%%%%%%%%%%%%%%%%%%%%%%%%%%%%%
%%%%%%%%%%%%%%%%%%%%%%%%%%%%%%%
\subsection{The ``quantum soliton''}
%%%%%%%%%%%%%%%%%%%%%%%%%%%%%%%
%%%%%%%%%%%%%%%%%%%%%%%%%%%%%%%
%%%%%%%%%%%%%%%%%%%%%%%%%%%%%%%
%%%%%%%%%%%%%%%%%%%%%%%%%%%%%%%
%%%%%%%%%%%%%%%%%%%%%%%%%%%%%%%
%%%%%%%%%%%%%%%%%%%%%%%%%%%%%%%%%%%%
%%%%%%%%%%%%%%%%%%%%%%%%%%%%%%%%%%%%
In the nonlinear case the ``quantum dynamics''
is governed by $\hat{H}=\hat{H}_0+\hat{H}_1$
in Eq. (\ref{qh}). Introducing the Bose field $\hat{\psi}$ in 
configuration space,
%

%{\em fiop}
\begin{equation}
\hat{\psi}(x) = (1/\sqrt{L})\sum_ka_k\exp{(ikx)} 
\label{fiop}
\end{equation}
the Hamiltonian  takes the form
%

%{\em h10}
\begin{equation}
\hat{H} = (-i)(\Delta/\nu)\int dx|\nabla\hat{\psi}^\dagger|^2 
-\lambda(\Delta/2\nu)^{3/2}\int dx(\hat{\psi}^\dagger-
\hat{\psi})^2\nabla(\hat{\psi}^\dagger+\hat{\psi})
\label{h10}
\end{equation}
describing the many-body interaction 
between the linear
diffusive modes governed by the first term $\hat{H}_0$. 

Imposing the  constraint of a horizontal interface 
we obtain
$\langle\hat{u}\rangle\propto\langle\hat{\psi}-\hat{\psi}^\dag\rangle=0$,
which implies that
$\langle\hat{\psi}\rangle = \langle\hat{\psi}^\dag\rangle$.
Since the interaction term $\hat{H}_1$ does not conserve the number
of particles, this constraint can only be satisfied for non-vanishing
$\langle\hat{\psi}\rangle$ if the diffusive
modes condense into  a {\em coherent condensate} so that
$\langle\hat{\psi}\rangle = \langle\hat{\psi}^\dag\rangle\neq 0$.
The resulting macroscopic wave function or condensate corresponds to
the ``classical'' soliton
mode discussed in the previous sections. The situation is
quite similar to the phenomenological theory for superfluid 
Helium based on a condensate wave function.
The condensate has two components, $\langle\hat{\psi}\rangle$
and $\langle\hat{\psi}^\dag\rangle$ or $\langle\hat{u}\rangle = u$
and $\langle\hat{\varphi}\rangle = \varphi$, and satisfies the coupled
field equations (\ref{eq1}-\ref{eq2}), obtained from the 
``quantum field equations''
(\ref{clf1}-\ref{clf2}) by ignoring ``quantum fluctuations'' 
and replacing the terms,
$\lambda\hat{u}\nabla\hat{u}$ and $\lambda\hat{u}\nabla\hat{\varphi}$,
by their average values, $\lambda u\nabla u$ and $\lambda u\nabla\varphi$.
We can thus regard Eqs. (\ref{eq1}-\ref{eq2}) as two coupled 
Gross-Pitaevsky-type
equations for the condensate wave function or soliton mode
\cite{pathria}.

The ``classical'' soliton is localized in space and carries energy  and 
momentum, depending on the boundary conditions according to the 
expressions (\ref{e}-\ref{p}). Subject to ``quantization'' this mode becomes a 
{\em bona fide }``quantum mechanical'' quasi-particle with the 
same energy and momentum.
Notice, however, that the ``quantum soliton'' is delocalized owing to
the ``uncertainty principle'' which implies that 
$\Delta x_0\Delta P\sim\Delta/\nu$; here $\Delta x_0$ is the uncertainty
in the center of mass position for the soliton and $\Delta P$ the uncertainty
in its momentum. For a ``quantum soliton'' with well-defined momentum
$P$ and energy $E$ we can in the usual way associate a wave number
$K$ and a frequency $\Omega$, according to the ``de Broglie'' relations,
$P=(\Delta/\nu)K$ and $E=(\Delta/\nu)\Omega$, and describe the 
quasi-particle by means of the wave function 
$\Psi\propto \exp{[-i\Omega t+iKx]}$.
Considering in particular a pair of ``quantum solitons'', describing a
propagating step in the height profile with energy and momentum given
by Eqs. (\ref{eee}-\ref{ppp}), we obtain
%

%{\em f1,w1}
\begin{eqnarray}
\Omega_{step}=&&i\left(\frac{\nu}{\Delta}\right) 
\frac{8}{3}\frac{|v|^3}{\lambda^2}
\label{f1}
\\
K_{step}=&&-i\left(\frac{\nu}{\Delta}\right) 4\frac{v|v|}{\lambda^2} ~ .
\label{w1}
\end{eqnarray}
and the wave function takes the form

%{\em wf1}
%
\begin{equation}
\Psi\propto\exp{[-i\Omega t+iKx]}=\exp{[\mbox{const}(x-v_{ph}t)]} ~ ,
\label{wf1}
\end{equation}
corresponding to a propagation with phase velocity  $v_{ph}=(2/3)v$. 
Noting, however, that the appropriate wave function for a
localized soliton is the wave packet construction,

%{\em wf2}
%
\begin{equation}
\Psi_{WP}\propto\sum_KA_K\exp{[-i\Omega t+iKx]}
\label{wf2}
\end{equation}
obtained  from a superposition of plane waves, we obtain the group velocity

%{\em vg}
%
\begin{equation}
v_g=\frac{d\Omega}{dK}=\frac{d\Omega/dv}{dK/dv} = v ~ .
\label{vg}
\end{equation}
This shows that the quasi-particle wave packet propagates with 
the same velocity
as the ``classical'' soliton in complete accordance with ``the 
correspondence principle''.
Whereas the propagation velocity $v$ determines the kinetics of
the ``classical'' soliton, the energy and momentum are the fundamental
characteristics in the ``quantum'' case; the velocity $v$ becoming the
group velocity of the wave packet. We also notice from the wave packet
form in Eq. (\ref{wf2}) that the ``quantum soliton'' corresponds to a
propagating mode. Finally, eliminating the velocity from 
Eqs. (\ref{f1}-\ref{w1})
we derive the ``quantum soliton'' dispersion law

%{\em d11}
%
\begin{equation} 
\Omega_K= i\lambda\frac{1}{3}\left(\frac{\Delta}{\nu}\right)^\frac{1}{2}
|K|^\frac{3}{2}
\label{d11}
\end{equation}
where $\text{sign}(\text{Im}K)=-\text{sign}v$.
%%%%%%%%%%%%%%%%%%%%%%%%%%%%%%
%%%%%%%%%%%%%%%%%%%%%%%%%%%%%%
%%%%%%%%%%%%%%%%%%%%%%%%%%%%%%
%%%%%%%%%%%%%%%%%%%%%%%%%%%%%%
%%%%%%%%%%%%%%%%%%%%%%%%%%%%%%
%%%%%%%%%%%%%%%%%%%%%%%%%%%%%%
%%%%%%%%%%%%%%%%%%%%%%%%%%%%%%
%%%%%%%%%%%%%%%%%%%%%%%%%%%%%%
\subsection{``Quantum fluctuations''}
%%%%%%%%%%%%%%%%%%%%%%%%%%%%%%
%%%%%%%%%%%%%%%%%%%%%%%%%%%%%%
%%%%%%%%%%%%%%%%%%%%%%%%%%%%%%
%%%%%%%%%%%%%%%%%%%%%%%%%%%%%%
%%%%%%%%%%%%%%%%%%%%%%%%%%%%%%
%%%%%%%%%%%%%%%%%%%%%%%%%%%%%%
%%%%%%%%%%%%%%%%%%%%%%%%%%%%%%
The final issue to consider in the qualitative ``quantization'' of the
soliton system is the role of ``quantum fluctuations'' in the presence
of a ``quantum soliton''. This problem is treated here by expanding the
fields $\hat{u}$ and $\hat{\varphi}$ about a soliton or condensate
configuration ($u_0,\varphi_0$). Inserting $\hat{u}=u_0+\delta\hat{u}$
and $\hat{\varphi}=\varphi_0+\delta\hat{\varphi}$ in 
Eqs. (\ref{clf1}) and (\ref{clf2}) we obtain to linear order  two coupled
equations for the ``quantum fluctuations'' $\delta\hat{u}$ and
$\delta\hat{\varphi}$,

%{\em fluc1, fluc2}
%
\begin{eqnarray}
\frac{\partial\delta\hat{u}}{\partial t} =&&
-i\nu\nabla^2\delta\hat{\varphi}+\lambda u_0\nabla\delta\hat{u}
+\lambda(\nabla u_0)\delta\hat{\varphi}
\label{fluc1}
\\
\frac{\partial\delta\hat{\varphi}}{\partial t} =&&
+i\nu\nabla^2\delta\hat{u}+\lambda u_0\nabla\delta\hat{\varphi}
+\lambda(\nabla \varphi_0)\delta\hat{u} ~ .
\label{fluc2}
\end{eqnarray}
These equations have the same form as the ones obtained
by expanding in the Gaussian fluctuations about the stationary soliton
solution in the path integral. 

The equations of motion (\ref{fluc1}-\ref{fluc2}) describe the interaction
of the linear ``quantum diffusive modes'' ($\delta\hat{u},\delta\hat{\varphi}$)
with the soliton configuration ($u_0,\varphi_0$) and constitute a
generalization to the noisy case of the linear stability equation in the
analysis in A; the soliton again acts like a potential giving rise
to phase shift effects and a gap in the diffusive spectrum.

Like in the noiseless case, the equations (\ref{fluc1}-\ref{fluc2}) admit 
an analytical solution.
Since the equations are Galilean invariant we need only consider the case
of a static soliton. 
First noting that the soliton solution according to Eq. (\ref{tsol}) is 
confined to the diagonal lines
$\nabla\varphi_0=i\mu\nabla u_0, \mu=\pm 1$, in phase space,
the fluctuations $\delta\hat{u}$,
$\delta\hat{\varphi}$ are disentangled by transforming to 
``normal coordinates'' along and perpendicular to the
``soliton lines''. Thus 1) introducing ``normal coordinates''
$\delta\hat{X}=\delta\hat{u}+i\delta\hat{\varphi}$,
$\delta\hat{Y}=\delta\hat{u}-i\delta\hat{\varphi}$,
2) inserting the static solution (\ref{stsol}), $u_0=\mu|u_+|\tanh{(k_sx)},
\nabla u_0=\mu|u_+|k_s\cosh^{-2}{(k_sx)}$, $\nabla\varphi_0=\mu\nabla u_0$,
and 3) performing the scaling transformations
$\delta\hat{X}\rightarrow h\delta\hat{X}$,
$\delta\hat{Y}\rightarrow h^{-1}\delta\hat{X}$, where $h=\cosh^\mu{(k_sx)}$,
in order to absorb the linear terms in $\nabla$, i.e.,
%{x,y,h,k}
%
\begin{eqnarray}
\delta\hat{X} = &&h^{-1}(\delta\hat{u}+i\delta\hat{\varphi})
\label{x}
\\
\delta\hat{Y} = &&h(\delta\hat{u}-i\delta\hat{\varphi})
\label{y}
\\
h=&&\cosh^\mu{(k_sx)}
\label{h}
\\ 
k_s=&&\frac{\lambda|u_+|}{2\nu}
\label{k}
\end{eqnarray}
we arrive at the effectively decoupled equations for the ``normal
coordinates''
\begin{eqnarray}
\frac{\partial\delta\hat{X}}{\partial t} =&&
+D\delta\hat{X}+(\mu-1)\nu k_s^2\delta\hat{Y}
\label{h1}
\\
\frac{\partial\delta\hat{Y}}{\partial t} =&&
-D\delta\hat{X}+(\mu+1)\nu k_s^2\delta\hat{Y}  ~ .
\label{h2}
\end{eqnarray}
Here the Schr\"{o}dinger operator $D$ has the same form as the stability
matrix for the noiseless Burgers equation in A or the sine Gordon
equation \cite{scott,fogedby5},

%{\em D}
%
\begin{equation}
D = -\nu\nabla^2+\nu k_s^2[1-\frac{2}{\cosh^2{(k_sx)}}] ~ .
\label{D}
\end{equation}
The wave number $k_s=\lambda|u_+|/2\nu$ introduced in section III 
depending on the soliton
amplitude sets the inverse length scale.
We also note that Eqs. (\ref{h1}-\ref{h2}) reduce to the linear case 
for $\lambda=0$
since $D\rightarrow-\nu\nabla^2$ and $h\rightarrow 1$.

Since the Bargman potential $\cosh^{-2}{(k_sx)}$ admits an exact solution
the spectrum of $D$ defined by the eigenvalue equation 
$D\Psi_n=i\omega_n\Psi_n$ is well-known \cite{landau2} and is discussed in A.
It is composed of a zero-eigenvalue localized bound state mode and a band of 
phase shifted scattering modes,

%{\em st11, st21}
%
\begin{eqnarray}
\Psi_0\propto&&\frac{1}{\cosh{(k_sx)}} 
\label{st11}
\\
\omega_0 =&& 0
\label{st11a}
\\
\Psi_k\propto&&\exp{(ikx)}\frac{k+ik_s\tanh{(k_sx)}}{k-ik_s} 
\label{st21}
\\
\omega_k=&&-i\nu( k^2+k_s^2) ~ .
\label{st21a}
\end{eqnarray}
Expanding $\delta\hat{X}$ and $\delta\hat{Y}$ on a set of eigenstates
$\Psi_n$

%{\em st3, st4}
%
\begin{eqnarray}
\delta\hat{X} =&& \sum_n\hat{a}_n\Psi_n
\label{st3}
\\
\delta\hat{Y} =&& \sum_n\hat{b}_n\Psi_n
\label{st4}
\end{eqnarray}
we finally obtain equations of motion for the expansion coefficients
$a_n$ and $b_n$,

%{\em A, B}
%
\begin{eqnarray}
\frac{d\hat{a}_n}{dt} =&& +i\omega_n\hat{a}_n+(\mu-1)\nu k_s^2\hat{b}_n
\label{A}
\\
\frac{d\hat{b}_n}{dt} =&& -i\omega_n\hat{b}_n+(\mu+1)\nu k_s^2\hat{a}_n
\label{B}
\end{eqnarray}
which we proceed to discuss.
%%%%%%%%%%%%%%%%%%%%%
%%%%%%%%%%%%%%%%%%%%%
%%%%%%%%%%%%%%%%%%%%%
%%%%%%%%%%%%%%%%%%%%%
%%%%%%%%%%%%%%%%%%%%%
%%%%%%%%%%%%%%%%%%%%%
%%%%%%%%%%%%%%%%%%%%%
\subsubsection{The translation modes}
%%%%%%%%%%%%%%%%%%%%%
%%%%%%%%%%%%%%%%%%%%%
%%%%%%%%%%%%%%%%%%%%%
%%%%%%%%%%%%%%%%%%%%%
%%%%%%%%%%%%%%%%%%%%%
%%%%%%%%%%%%%%%%%%%%%
%%%%%%%%%%%%%%%%%%%%%
The zero-frequency of the Schr\"{o}dinger operator $D$ in
Eq. (\ref{D}) is associated with the translation and boosting of the
static soliton profile $(u_0,\varphi_0)$. This is seen in the following way.
Since the ``quantum field equations'' (\ref{clf1}-\ref{clf2}) have 
the same form 
as the 
``classical''
field equations (\ref{eq1}-\ref{eq2}) they are equally satisfied by a
soliton solution. Consequently, a variation of the static soliton
profile, $(\delta u_0,\delta\varphi_0)$, is a solution of the linearized
equations
(\ref{fluc1}-\ref{fluc2}) or (\ref{h1}-\ref{h2}) in the static case,
corresponding to the bound state $\Psi_0=0$. 
Furthermore, since the 
soliton depends 
parametrically on
the center of mass position $x_0$ it follows that the fluctuations
$(\delta u_0,\delta\varphi_0)$ are proportional to the derivatives
$(\nabla u_0,\nabla\varphi_0)$ with respect to $x_0$, corresponding
to a displacement of the soliton position.
This mode
is thus a translation or Goldstone mode associated with the broken 
translational symmetry and is a well-known feature of symmetry breaking
``excitations''; a Goldstone mode is excited in order to restore the 
broken symmetry 
\cite{chaikin}. 
A similar translation mode was 
also encountered in
our discussion of the noiseless case in A.

Focussing for example on the {\em right hand} static soliton for 
$\mu=+1$ and solving 
Eqs. (\ref{A}-\ref{B}) for $n=0$ we have the expansion coefficients
\begin{eqnarray}
\hat{a}_0 =&& \text{cst.}
\label{c}
\\
\hat{b}_0 =&&\hat{b}_0^0 + 2\nu k_s^2 \hat{a}_0 t ~ ,
\label{d}
\end{eqnarray}
where $\hat{b}_0^0$ is the initial value for $t=0$, and we obtain, using that
$\nabla u_0\propto\cosh^{-2}(k_sx)$, the fluctuation mode
\begin{equation}
\delta\hat{u}=\hat{a}_0 + (\hat{b}_0^0 + \hat{a}_0\nu k_s^2 t)\nabla u_0 ~ .
\end{equation}
For $\hat{a}_0 = 0$ this mode corresponds to an infinitesimal translation
$\delta x_0=\hat{b}_0^0$ of the soliton, i.e., a change of the center of mass 
coordinate; for $\hat{a}_0\neq 0$ the mode is equivalent to a boost of the 
soliton to a small velocity $\propto\nu k_s$. 
A similar discussion applies to $\delta\hat{\varphi}$.
%%%%%%%%%%%%%%%%%%%%%
%%%%%%%%%%%%%%%%%%%%%
%%%%%%%%%%%%%%%%%%%%%
%%%%%%%%%%%%%%%%%%%%%
%%%%%%%%%%%%%%%%%%%%%
%%%%%%%%%%%%%%%%%%%%%
%%%%%%%%%%%%%%%%%%%%%
\subsubsection{The diffusive scattering modes}
%%%%%%%%%%%%%%%%%%%%%
%%%%%%%%%%%%%%%%%%%%%
%%%%%%%%%%%%%%%%%%%%%
%%%%%%%%%%%%%%%%%%%%%
%%%%%%%%%%%%%%%%%%%%%
%%%%%%%%%%%%%%%%%%%%%
%%%%%%%%%%%%%%%%%%%%%
The band of diffusive scattering modes are also easily discussed.
From the equations of motion for the expansion coefficients
in Eqs. (\ref{A}-\ref{B}) and again considering a {\em right hand} 
static soliton for $\mu=1$ we obtain for $n=k$ the solution
\begin{eqnarray}
\hat{a}_k =&& \hat{a}_k^0 e^{i\omega_kt} 
\\
\hat{b}_k =&&\hat{a}_k^0\left[\frac{k_s^2}{k^2+k_s^2}\right]e^{i\omega_kt} +
\left[\hat{b}^0_k - 
\hat{a}^0_k\left[\frac{k_s^2}{k^2+k_s^2}\right]\right]e^{-i\omega_kt} ~ .
\end{eqnarray}
Here $a_k^0$ and $b_k^0$ are the initial values and the spectrum
$\omega_k$ given by Eq. (\ref{st21a}). For the fluctuation $\delta\hat{u}$
we then have
\begin{equation}
\delta\hat{u} = \sum_k[\hat{a}_k\cosh{(k_sx)}+
\hat{b}_k\cosh^{-1}{(k_sx)}]\Psi_k ~ ,
\end{equation}
where $\Psi_k$ is given by Eq. (\ref{st21}). We note that $\delta\hat{u}$
in the soliton case again is composed of both positive and negative
frequency parts,
$\exp{(\nu(k^2+k_s^2)t)}$
and $\exp{(-\nu(k^2+k_s^2)t)}$, corresponding to growing and decaying
modes in the stationary driven regime, exhibiting a gap $\nu k^2_s$ in the 
spectrum. In the linear EW case $k_s=0$ and $\delta\hat{u}$ assumes the
form in Eq. (\ref{opu}).
We shall not here dwell on the somewhat complicated $x$-dependence but 
only observe that the main effect of the soliton on the diffusive modes
apart from phase shift effects and spatial modulations is to lift the
spectrum and create a gap $\nu k^2_s=\lambda^2|u_+|^2/4\nu$
depending  on the soliton amplitude
$u_+$, the coupling $\lambda$ and the frequency $\nu$.
%%%%%%%%%%%%%%%%%%%%%
%%%%%%%%%%%%%%%%%%%%%
%%%%%%%%%%%%%%%%%%%%%
%%%%%%%%%%%%%%%%%%%%%
%%%%%%%%%%%%%%%%%%%%%
%%%%%%%%%%%%%%%%%%%%%
%%%%%%%%%%%%%%%%%%%%%
\subsection{Many-body description of a growing interface}
%%%%%%%%%%%%%%%%%%%%%%%%%%%%%%%%
%%%%%%%%%%%%%%%%%%%%%%%%%%%%%%%%
%%%%%%%%%%%%%%%%%%%%%%%%%%%%%%%
%%%%%%%%%%%%%%%%%%%%%%%%%%%%%%%%
%%%%%%%%%%%%%%%%%%%%%%%%%%%%%%%%
%%%%%%%%%%%%%%%%%%%%%%%%%%%%%%%%
The above discussion of the ``quantum solitons'' and the 
``quantum diffusive modes'' allow a heuristic qualitative discussion
of a growing interface. The stochastic dynamics of the noisy
Burgers equation (\ref{bur2}) in the stationary regime can be
rigorously interpreted in terms of a dilute Landau-type ``quantum''
quasi-particle gas composes of elementary excitations of two types:
``Quantum solitons'' and ``quantum diffusive modes''. 
The ``quantum mechanics'' being equivalent to a Master equation
description \cite{fogedby3,stinchcombe} is basically relaxational
corresponding to a complex Hamiltonian.

The elementary excitations fall in two classes: Linear diffusive modes
and nonlinear soliton modes. 1) The linear diffusive modes are
are associated with the damping term in the Burgers equation or,
equivalently, the harmonic anti-Hermitian part in the Hamiltonian.
These modes account for the relaxational aspects of the interface and are
characterized by the dispersion law (\ref{st21a}), i.e.,
\begin{equation}
\omega_k = -i\nu(k^2+k_s^2)
\label{DL}
\end{equation}
with a gap $\nu k_s^2$. As in our discussion of the noiseless case
in A the gap can be associated with a non-vanishing current towards
the center of the soliton where the damping in enhanced.
2) The nonlinear soliton modes are related to the nonlinear growth term
in the Burgers equation or, equivalently, the nonlinear Hermitian
part of the Hamiltonian. For a pair of solitons representing a
growing step the dispersion law is given by Eq.(\ref{d11}), i.e., 
\begin{equation}
\Omega_K = i\frac{\lambda}{3}\left(\frac{\Delta}{\nu}\right)^\frac{1}{2}
|K^|\frac{3}{2} ~ .
\label{Q}
\end{equation}
The mode is gapless and characterized by the fractional exponent $3/2$.
The soliton mode accounts for the growth aspects of the driven interface.
For $\nu\rightarrow\infty$ the linear damping dominates  the growth
and $\Omega_K\rightarrow 0$, also for $\lambda\rightarrow 0$ we attain the
linear EW case and  $\Omega_K\rightarrow 0$; finally for $\Delta\rightarrow 0$
the stochastic aspects are quenched, solitons (and diffusive modes) are not
kinetically or stochastically excited  and $\Omega_K\rightarrow 0$.

In the linear EW case the fluctuating interface is in equilibrium
and here described as a non-interacting gas of linear gapless
diffusive modes. The statistical fluctuations appear as ``quantum
fluctuations'' of the quasi-particle modes. Since the dispersion is 
quadratic we can also envisage the EW case as a ``quantum'' gas of
free particles with imaginary mass.

In the nonlinear Burgers case the ``quantum soliton'' emerges as
a new additional quasi-particle, corresponding to the faceted genuine
growth of an interface. The linear modes become subdominant in the sense
that they develop a gap in the spectrum and correspond to superposed
damped ``ripple modes'' on the soliton configurations. The diffusive modes
extend over the whole configuration and are phase-shifted due to 
reflectionless scattering off the solitons like in the noiseless case.
A scattering analysis also, in accordance with Levinson's theorem, shows
the diffusive spectrum is depleted by a number of states, corresponding 
to the translation modes of the solitons.

Before turning to the heuristic scaling analysis in the next section,
we wish to add a few more remarks concerning the structure of
a field theoretic or many-body description of the noisy Burgers equation.
There are basically two equivalent modes of approach:
1) A direct evaluation of the path integral in the weak noise limit
in a saddle point approximation for a dilute soliton gas, including
the diffusive modes, corresponding to Gaussian fluctuations about
the saddle points, and summing over periodic orbits in order to
include secular effects or
2) A construction of the equivalent ``quantum many-body theory'' on
the basis of the ``quantum representation'' of the path integral.
Including the soliton modes as space- and time-dependent {\em condensate}
configurationsi, as mentioned in section XB, the many-body approach is
similar to the microscopic theory on interacting bosons 
\cite{mahan,abrikosov,landau3} with anomalous propagators etc. There are,
however, some notable differences. In the case of interacting bosons
the uniform condensate acts as a particle reservoir and changes the free
boson dispersion law $\omega\propto p^2$ to a linear acoustic phonon
branch $\omega\propto p$. In the present case , the condensate corresponding
to a soliton or gas of solitons is non-uniform and time-dependent,
governed by the ``classical'' field equations. The free diffusive modes
with dispersion $\omega\propto p^2$ develop a gap depending on the soliton
amplitude or velocity, whereas the soliton or condensate mode emerges as
a new quasi-particle with dispersion $\omega\propto p^{3/2}$.
%%%%%%%%%%%%%%%%%%%%%%%%%%%%%%%%
%%%%%%%%%%%%%%%%%%%%%%%%%%%%%%%
%%%%%%%%%%%%%%%%%%%%%%%%%%%%%%%
%%%%%%%%%%%%%%%%%%%%%%%%%%%%%%%
%%%%%%%%%%%%%%%%%%%%%%%%%%%%%%%
%%%%%%%%%%%%%%%%%%%%%%%%%%%%%%%
%%%%%%%%%%%%%%%%%%%%%%%%%%%%%%%%
\section{Scaling and Universality classes}
%%%%%%%%%%%%%%%%%%%%%%%%%%%%%%%
%%%%%%%%%%%%%%%%%%%%%%%%%%%%%%%
%%%%%%%%%%%%%%%%%%%%%%%%%%%%%%%
%%%%%%%%%%%%%%%%%%%%%%%%%%%%%%%
%%%%%%%%%%%%%%%%%%%%%%%%%%%%%%%
%%%%%%%%%%%%%%%%%%%%%%%%%%%%%%%
%%%%%%%%%%%%%%%%%%%%%%%%%%%%%%%
In addition to providing
a many-body description of a growing interface in terms of a Landau-type
quantum quasi-particle gas of propagating ``quantum solitons'' and 
damped ``quantum diffusive modes'', the path integral formulation also
offers as a by-product some insight into the scaling properties,
i.e., the behavior of the interface in the limit of large distances
and long times.

We shall here focus on the scaling properties of the slope correlation
function summarized in the dynamical scaling form (\ref{sf}),
i.e., assuming $t=0$ in the stationary regime,
%{\em 1}
%
\begin{equation}
\langle u(x,t)u(0,0)\rangle = |x|^{2(\zeta - 1)}f(|t|/|x|^z)
\label{1}
\end{equation}
The scaling issue is then to determine 
the roughness or wandering
exponent $\zeta$, the dynamic exponent $z$, and the scaling function
$f(w)$. 

In the EW case the scaling function $f$ is given by (\ref{sfew}), i.e., 
%{\em 2}
%
\begin{equation}
f(w)=(\Delta/2\nu)(4\pi\nu)^{-1/2}w^{-1/2}\exp{[-1/4\nu w]} ~ ,
\label{2}
\end{equation}
implying the exponents $(\zeta,z)=(1/2,2)$. In the Burgers case Galilean 
invariance leads to the scaling law (\ref{sl}), i.e.,
$\zeta+z=2$, which together with the stationary distribution (\ref{gauss}),
an effective fluctuation-dissipation theorem, yields the exponents
$(\zeta,z)=(1/2,3/2)$.

According to the path integral formulation in sections IV and V, using
Eqs. (\ref{cor}) and (\ref{gen3}) the slope correlation function is given by
%{\em 3}
%
\begin{equation}
\langle u(x,t)u(0,0)\rangle =
Z(0)^{-1}
\int\prod_{xt} dud\varphi\exp{\left[i\frac{\nu}{\Delta}S\right]}u(xt)u(00) ~ ,
\label{3}
\end{equation}
or in terms of the underlying ``quantum field theory'', noting that the 
path integral by construction defines time-ordered products \cite{das},
%{\em 4}
%
\begin{equation}
\langle u(x,t)u(0,0)\rangle =
\langle 0|T\hat{u}(x,t)\hat{u}(0,0)|0\rangle ~ .
\label{4}
\end{equation}
Here $|0\rangle$ denotes the appropriate stationary state for the
system \cite{footnote8}.

In order to elucidate the structure of Eq. (\ref{4}) we construct
a spectral representation by i) displacing the slope field
$\hat{u}(x,t)$ to the origin in space and time by 
means of the Hamiltonian $\hat{H}$ and the momentum $\hat{P}$, using
the integrated form of the commutator relations (\ref{heis1}) and
(\ref{comm1}), and ii) inserting intermediate eigenstates $|P\rangle$
with momentum $P$ and energy $E_P$. The first step implies the relation
\begin{equation}
\hat{u}(x,t) = 
\exp{\left[i\frac{\nu}{\Delta}(\hat{P}x+\hat{H}t)\right]}
\hat{u}(0,0)
\exp{\left[-i\frac{\nu}{\Delta}(\hat{P}x+\hat{H}t)\right]} ~ ;
\label{5}
\end{equation}
secondly, inserting intermediate states, using 
$\hat{H}|P\rangle=E_P|P\rangle$ and $\hat{P}|P\rangle=P|P\rangle$,
introducing the frequency and wave number $(\Omega_K,K)$ associated with
the elementary excitations or quasi-particles, and lumping the matrix elements
in an effective form factor, 
$G(K)=\langle 0|\hat{u}|K\rangle\langle K|\hat{u}|0\rangle$, we arrive at the
spectral representation
\begin{equation}
\langle u(x,t)u(0,0)\rangle = 
\int\frac{dK}{2\pi}G(K)\exp{[-i(\Omega_K|t|-iK|x|)]} ~ .
\label{6}
\end{equation}
The time-ordering in Eq. (\ref{4}) together with parity invariance,
$x\rightarrow-x$, imply evenness in the dependence on $x$ and $t$;
also $G(K)$ must be even in $K$.

The spectral form (\ref{6}) is only schematic. For a multi-soliton
diffusive mode intermediate eigenstate $|\{K_i\},\{k_j\}\rangle$,
where $K_i$ and $k_j$ denote the soliton and diffusive mode wave numbers,
respectively, with total wave number $K=\sum_iK_i+\sum_jk_j$
and total frequency $\Omega=\sum_i\Omega_{K_i}+\sum_j\omega_{k_j}$,
we have strictly speaking the spectral form, say for $t>0$,
\begin{equation}
\langle u(x,t)u(0,0)\rangle =
\int\prod_{ij}dK_idk_jG(\{K_i\},\{k_j\})
e^{-i[(\sum_iK_i+\sum_jk_j)x+(\sum_i\Omega_{K_i}+\sum_j\omega_{k_j})t]} ~ .
\label{7}
\end{equation}

Since the soliton are transparent with respect to the diffusive modes
as discussed in section XC,
the operator $\hat{u}$ only excites a single mode $k$ 
extending across the system,
i.e., $G(\{K_i\},\{k_j\})\sim G(\{K_i\},k)$, and assuming furthermore
that $G(\{K_i\},k)$ factorizes approximately in accordance with
the dilute soliton gas picture, $G(\{K_i\},k)\sim G_D(k)\prod_iG_S(K_i)$,
we obtain summing over the solitons
\begin{equation}
\langle u(x,t)u(0,0)\rangle\sim
\frac{\int dkG_D(k)e^{-ikx-i\omega t}}{1-\int dKG_S(K)e^{-iKx-i\Omega t}} ~ .
\label{8}
\end{equation}
The expression (\ref{8}) is clearly not correct in detail since 
we have not solved
the many-body problem but only made some plausible assumptions 
concerning the form factor $G$. Nevertheless, from the point of view
of discussing the scaling properties Eq. (\ref{8}) has the required
structure and serves our purpose. In the EW case $G_S(K)=0$ and 
Eq. (\ref{8}) reduces to the scaling form  (\ref{2}). In the Burgers
case $\Omega\propto i|K|^{3/2}$, i.e., 
$\exp{(-i\Omega t)}= \exp{(+\text{const.}|K|^{3/2}t)}$, and the denominator
$(1-\int dKG_S(K)e^{-iKx-i\Omega t})^{-1}
\sim \int dKG_S(K)e^{-iKx-i\Omega t}$ controls the scaling behavior.
In both cases we can use the simplified general spectral form (\ref{6}).

For the purpose of a discussion of the scaling properties we first
consider a general quasi-particle dispersion law with a gap 
$\tilde{\Delta}$, stiffness constant $A$, and exponent $\beta$,
\begin{equation}
\Omega_K = \tilde{\Delta} + A|K|^\beta ~ .
\label{9}
\end{equation}
For large distances $|x|\gg a$, where $a$ is a microscopic length
defining the UV cut-off $K\sim 1/a$ implied in Eq. (\ref{6}), the 
spectral representation samples the small wave number regions 
$K\ll 1/a$. Assuming that the form factor is regular for small $K$,
i.e., $G(K)\sim G(0) +(1/2)K^2G''(0)$, and rescaling $K$, $Kx\rightarrow K$,
we obtain, inserting the general dispersion (\ref{9}), the spectral 
representation in a more appropriate scaling form
\begin{equation}
\langle u(x,t)u(0,0)\rangle\sim
G(0)e^{-i\tilde{\Delta}t}x^{-1}\int\frac{dK}{2\pi}
e^{-iA|K|^\beta |t|/|x|^\beta - iK} ~ .
\label{10}
\end{equation}

We emphasize again that the spectral representation (\ref{10}) 
can only be considered
as a heuristic expression since we have not here carried out a detailed
analysis of the non-Hermitian non-Lagrangian field theory underlying the
path integral. Nevertheless, we believe that we can already here draw some
interesting general consequences concerning the scaling properties of
a growing interface.

First we observe that in the presence of a gap $\tilde{\Delta}\neq 0$ there
is no scaling behavior. For the diffusive mode in the presence of
a soliton the spectrum is for example given by Eq. (\ref{DL}) with
a gap $\tilde{\Delta} = -i\nu k_s^2$, implying an exponential fall-off with
$t$ in Eq. (\ref{10}). Consequently, only gapless excitations for
$\tilde{\Delta} = 0$ contribute to the scaling behavior. The gapless
excitations are associated with the so-called zero temperature fixed
point behavior of the ``quantum field theory'' and determine the scaling
properties.

Furthermore, comparing the spectral form (\ref{10}) for $\tilde{\Delta} = 0$
with the scaling form (\ref{1}) we immediately identify the roughness
exponent $\zeta =1/2$ and the dynamic exponent $z=\beta$. We also
note that whereas the exponent $\zeta=1/2$ essentially follows from
a simple {\em regularity property} of the form factor $G(K)$ with leading
term $G(0)$ for small $K$, the exponent $z$ is tied to the exponent $\beta$
in the quasi-particle dispersion law.

In the linear EW case the diffusive gapless modes with dispersion law
(\ref{d1}), i.e., $\omega_k=-i\nu k^2$, exhaust the spectrum and we
obtain $\beta=z=2$, corresponding to the EW universality class in Table 1.
Also the spectral form yields the scaling function (\ref{2}) with
the identification $G(0) = \Delta/2\nu$.

In the nonlinear Burgers-KPZ case the soliton modes with 
gapless dispersion (\ref{Q}), i.e., 
$\Omega_K\propto(\Delta/\nu)^{1/2}\lambda|K|^{3/2}$,
exhaust the bottom of the spectrum and yields the exponent
$\beta = z =3/2$, corresponding to the Burgers-KPZ universality class
in Table 1;
the linear diffusive modes develop a gap according to Eq. (\ref{DL}),
become subdominant and do not contribute to the scaling behavior.

The above discussion thus provides a {\em dynamical interpretation} of the 
scaling properties, exponents, and universality classes. The universality 
class is determined by the lowest-lying gapless excitation. The spectral
form also elucidates the {\em robustness} of the roughness exponent 
$\zeta$ which is the same for both universality classes. In the case
of the stationary equal-time fluctuations we set $t=0$ in Eq. (\ref{8})
and the resulting scaling form yielding $\zeta$ does {\em not}
depend on the specific quasi-particle dispersion law; this argument is
equivalent to the effective fluctuation-dissipation theorem yielding
the stationary distribution (\ref{gauss}) {\em independent} of the
nonlinear drive $\lambda$.

The spectral form (\ref{10}) also provides an expression for the scaling
function $f(w)$. We obtain comparing Eq. (\ref{8}) with Eq. (\ref{1})
\begin{equation}
f(w) = G(0)\int\frac{dK}{2\pi}e^{-i(A|K|^zw + K)}
\label{11}
\end{equation}
The above expression is at best heuristic but we do notice that it has
the correct limiting behavior, i.e., $f(w)\sim\text{const}$ for
$w\rightarrow 0$ and $f(w)\sim w^{-1/z}$ for $w\rightarrow\infty$.

The scaling function $f(w)$ describing the behavior of the strong
coupling fixed point has been accessed both numerically 
\cite{krug5,tang,amar} and by means of an analytical mode coupling
approach \cite{frey2,hwa2}, based on a self-consistent one-loop
calculation, i.e., to first order in $\lambda$, and assuming vanishing
vertex corrections. The agreement between the numerical simulations
and the analytical method is good, indicating that the mode coupling
approach seems to capture essential properties of the strong coupling
fixed point behavior.

The heuristic and preliminary character of the scaling function
given here does not allow a detailed comparison. We note, however, 
that since $A\propto\lambda(\Delta/\nu)^{1/2}$ in the Burgers-KPZ
case the dimensionless argument in the scaling function $f(w)$ is
$\lambda((\Delta/\nu)^{1/2}t/x^{3/2}$. This is in complete agreement
with the driven lattice gas DRG analysis in \cite{janssen1} and with the
general arguments advanced in the mode coupling analysis in
\cite{frey2,hwa2}.

We also note the curious fact that the spectral form (\ref{6}) for a gapless
dispersion with exponent $\beta$ bears resemblance to the form of 
the probability
distribution for a one-dimensional L\'{e}vy flight with index
$\mu=\beta$ \cite{fogedby7}. The case $\mu=\beta=2$ corresponds to 
ordinary Brownian walk, whereas $\mu=\beta=3/2<2$ is equivalent to
super diffusion.

The present analysis of the scaling properties based on the spectral
form (\ref{6}) originates from a weak noise saddle point approximation to
the path integral and as such only holds for $\Delta\rightarrow 0$.
However, within the general assumptions underlying the application of
scaling theory and the notion of universality classes, we expect the
exponents and scaling function to be universal characteristics of the
system and thus {\em independent} of the noise strength $\Delta$.
This property can, however, be reconciled with the present many-body
approach if we assume that an enhancement of the noise strength, that
is a stronger drive of the system, only leads to a {\em dressing} of
the quasi-particle spectrum, i.e., a change in the stiffness constant,
and not to a change in the exponent $\beta$.
In the ``quantum mechanical'' language this corresponds to the assumption
that the WKB approximation also holds in the strong ``quantum regime''
as far as the exponent of the quasi-particle dispersion law is concerned.

We conclude this section with a few speculative remarks concerning
the ``breakdown of hydrodynamics''. The noisy Burgers equation is
basically a nonlinear conserved hydrodynamical equation derived
by combining  the conservation law $\partial u/\partial t +\nabla u=0$
with a constitutive equation for the current, 
$j=-\nu\nabla u -(\lambda/2)u^2-\eta$, with transport coefficients 
$\nu$ (the damping) and $\lambda$ (the mode coupling). The expression
for the deterministic part of $j$ is thus based on a gradient expansion
to lowest order and the simplest quadratic nonlinearity in $u$.
The issue is in which way the mode coupling term affects the hydrodynamical
properties. In this context ``breakdown of hydrodynamics'' usually
refers to the situation where the underlying regularity structure
of the gradient expansion, i.e., in wave number space regularity in an
expansion in $k$, breaks down. 

In the present many-body formulation, entailing the spectral form
(\ref{6}), we obtain in frequency space
\begin{equation}
\langle uu\rangle(k,\omega)\sim
\text{Re}\frac{1}{\omega - \Omega_k}
\end{equation}
In the linear EW case for $\omega_k^0=-i\nu k^2$ we recover the diffusive form
(\ref{lor}), corresponding to a diffusive pole $\omega_k^0 = -i\nu k^2$ in the
complex $\omega$ - plane. However, in the nonlinear mode coupling case for
$\lambda\neq 0$, $\Omega_k\propto|k|^{3/2}$ and $\langle uu\rangle(k,\omega)$
develops a branch cut structure, corresponding to a non-analytic
wave number dependence in the current, i.e., a breakdown of hydrodynamics.

%%%%%%%%%%%%%%%%%%%%%%%%%%%%%%%
%%%%%%%%%%%%%%%%%%%%%%%%%%%%%%%
%%%%%%%%%%%%%%%%%%%%%%%%%%%%%%%
%%%%%%%%%%%%%%%%%%%%%%%%%%%%%%%
%%%%%%%%%%%%%%%%%%%%%%%%%%%%%%%
%%%%%%%%%%%%%%%%%%%%%%%%%%%%%%%%
\section{Discussion and conclusion}
%%%%%%%%%%%%%%%%%%%%%%%%%%%%%%%
%%%%%%%%%%%%%%%%%%%%%%%%%%%%%%%
%%%%%%%%%%%%%%%%%%%%%%%%%%%%%%%
%%%%%%%%%%%%%%%%%%%%%%%%%%%%%%%
%%%%%%%%%%%%%%%%%%%%%%%%%%%%%%%
%%%%%%%%%%%%%%%%%%%%%%%%%%%%%%%
%%%%%%%%%%%%%%%%%%%%%%%%%%%%%%%
In the present paper we have advanced a novel approach to the growth
morphology and scaling behavior of the noisy Burgers equation in
one dimension. Using the Martin-Siggia-Rose (MSR) technique in a canonical
form we have demonstrated that the physics of the so far elusive 
strong coupling
fixed point  is associated with an essential singularity
in the noise strength and can be accessed by appropriate theoretical
soliton techniques.

The canonical representation of the MSR functional integral in terms
of a Feynman phase space path integral with a complex Hamiltonian
identifies the noise strength as the relevant small non-perturbative
parameter and allows for {\em a principle of least action}. In the
asymptotic weak noise limit the leading contributions to the path
integral are given by a dilute gas of solitons with superposed
linear diffusive modes. The canonical variables are the local slope
of the interface and an associated ``conjugate'' noise field, characteristic
of the MSR formalism. In terms of the local slope the soliton and diffusive
mode picture provide a many-body description of a growing interface governed
by the noisy Burgers equation. The noise-induced slope fluctuations
are here represented by the various paths or configurations contributing
to the path integral.

The canonical formulation of the path integral and the associated
principle of least action also allow us to associate energy, momentum, 
and action with a given soliton configuration or growth morphology.
This gives rise to {\em a dynamical selection criterion} similar
to the role of the Boltzmann factor $\exp{(-E/T)}$ in equilibrium
statistical mechanics which associates an energy $E$ with a given 
configuration contributing to the partition function; in the dynamical
case the action $S$ provides the weight function for the dynamical
configuration. More detailed, in the dynamical case, ``rotating'' the
noise variable, $\varphi\rightarrow -i\varphi$, we have the partition
function
\begin{eqnarray}
Z_{dyn}\propto&&\int\prod_{xt}dud\varphi\exp{[-\frac{\nu}{\Delta}\tilde{S}]}\\
\tilde{S}=&&\int dxdt[u\frac{\partial u}{\partial t} -\tilde{\cal H}
(u,\varphi)]\\
\tilde{\cal H} =&& -\frac{\nu}{2}[(\nabla u)^2-(\nabla\varphi)^2]+
\frac{\lambda}{2}u^2\nabla\varphi ~ ,
\end{eqnarray}
whereas in the equilibrium case we have the general form in 
one dimension,
\begin{eqnarray}
Z_{eq}\propto&&\int\prod_{x}dpdq\exp{[-\frac{1}{T}H]}\\
H=&&\int dx{\cal H}(p,q) ~ ,
\end{eqnarray}
where ${\cal H}$ is the Hamiltonian density. By comparison we
note that the noise strength $\Delta$ plays the role of a 
``temperature'' in the dynamical case. We also observe that
$Z_{dyn}$ for the dynamical 1-D problem, treating time as an additional
coordinate, i.e., $t\rightarrow y$, is equivalent to a 2-D equilibrium
partition function with Hamiltonian
\begin{equation}
H = \int dxdy[u\nabla_y\varphi + \frac{\nu}{2}
[(\nabla_x u)^2-(\nabla_x\varphi)^2]-\frac{\lambda}{2} u^2\nabla_x\varphi]
\end{equation}
and temperature $\nu/\Delta$.

In addition to providing a physical many-body picture of the morphology
of a growing interface in terms of soliton modes accounting for the
growth aspects and diffusive modes corresponding to the relaxational
aspects, the present approach also gives insight into the scaling 
properties. The perspective here is not a ``coarse-graining'' procedure,
replacing the original description by a scaling description with
ensuing dynamical renormalization group (DRG) equations, but rather 
a focus on the gapless elementary excitations or quasi-particles
of the many-body theory. 

The case of simple scaling characterized by a roughness exponent, a dynamical
exponent, and a scaling function, corresponding to a simple fixed point
structure in the DRG analysis, is here represented by a simple quasi-particle
mode exhausting the bottom of the spectrum with a gapless dispersion law.
The dynamic exponent is given by the exponent in the quasi-particle
dispersion law, whereas the roughness exponent follows from a regularity
property of the form factor in a spectral representation of the slope
correlations; the scaling function being given by the spectral
form itself.

Our analysis shows that the nonequilibrium growth dynamics in
one dimension is controlled by solitons or dynamic domain wall. In this
respect their is a parallel between the present kinetic growth problem
and other well-studied low dimensional equilibrium problems also
controlled by localized excitations such as the one dimensional
Ising model with domain wall excitations or the two dimensional
XY model characterized by vortex excitations \cite{chaikin}.
The present approach is conducted in one dimension and assumes a 
spatially short-range correlated conserved noise in order to
implement the shift transformation leading to the canonical formulation
and the separation of the Hamiltonian in a harmonic part and an interacting
part. In higher dimension the Burgers equation becomes a vector equation
with a nonlinear term $\lambda(\vec{u}\vec{\nabla})\vec{u}$ and we
obtain a more complicated Hamiltonian governing the dynamics. 
We have not pursued the higher dimensional case yet, but it is 
nevertheless interesting to speculate that the strong coupling
fixed point behavior in $d=2$ and below is associated with higher
dimensional localized soliton-like excitations, i.e., the dynamics
is defect-dominated.

Finally we wish to comment on some recent work on the driven Burgers
equation with noise at large length scales modelling forced turbulence.
This problem has been treated using a variety of methods such as
operator product expansions \cite{polyakov}, instanton calculations
\cite{gurarie}, and replica methods \cite{bouchaud}. In this context
the non-perturbative instanton methods used in order to
determine the tail of the velocity probability distribution seem
superficially to be related to the present soliton approach in that they are
also based on a saddle point approximation to the MSR functional 
integral. However, unlike the the soliton which is only localized
in space, the instanton is localized in both space and time,
yielding  a finite action, and does not represent a soliton in
the Burgers equation. The difference between the two approaches
is related to the spatial correlations of the noise driving the system:
In the forced turbulence case the noise is correlated at large distances,
whereas in the growth problem the noise is assumed to be conserved and
delta function correlated in space.
\acknowledgements
{
Discussions with J. Krug, M. Kosterlitz, M. H. Jensen, T. Bohr, M. Howard,
K. B. Lauritsen and A. Svane
are gratefully acknowledged.
}
%\vfill
%\eject
\newpage
\centerline{\bf\large REFERENCES}
%\twocolumn
%\narrowtext

%%%%%%%%%%%%%%%%%%%%%
%%%%%%%%%%%%%%%%%%%%%
\smallskip
\smallskip
\smallskip
\centerline{\bf\large FIGURES}
\begin{figure}
\centerline
{
\epsfxsize=10cm
\epsfbox{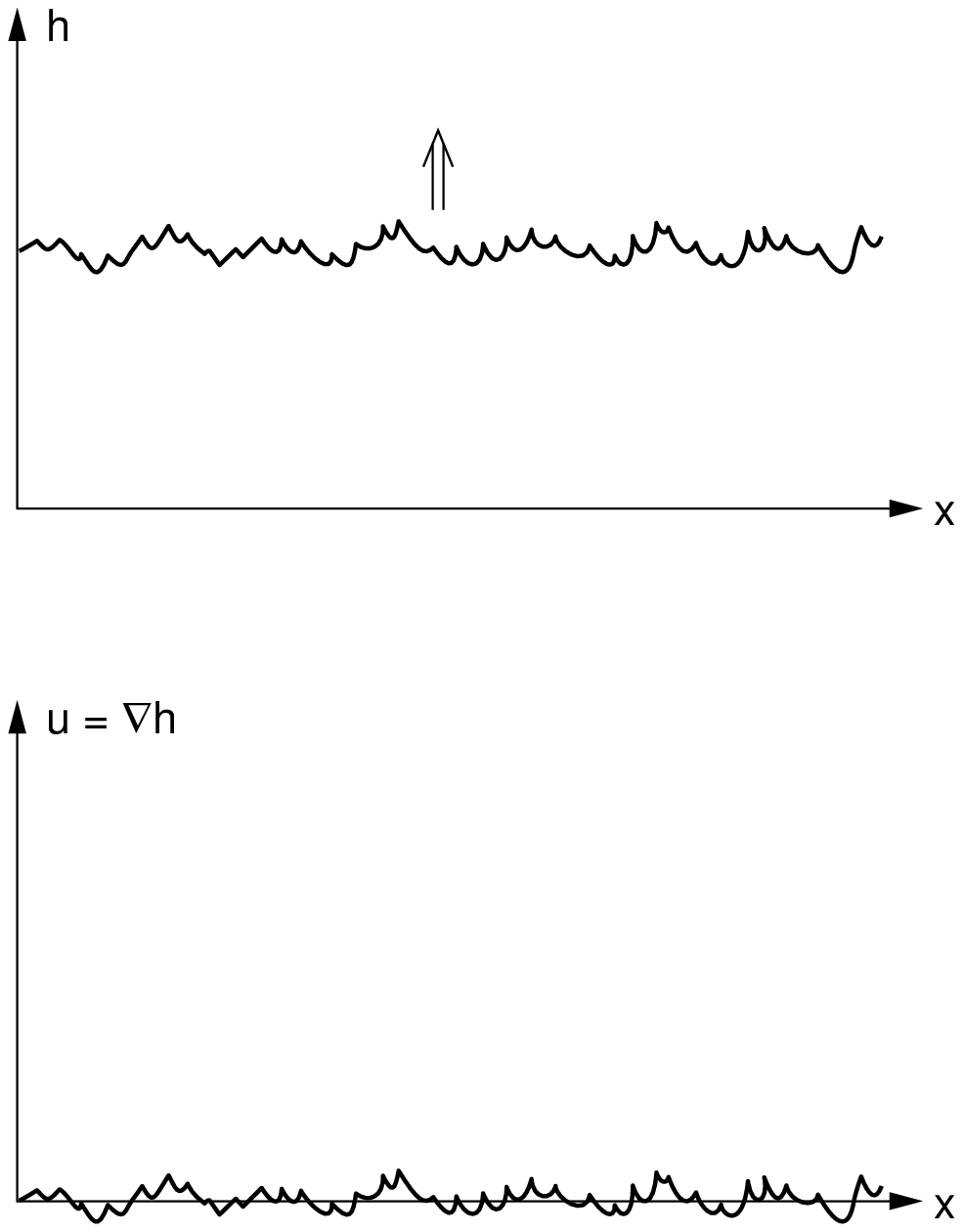}
}
\caption{
We depict the general growth morphology
for a 1-D interface in terms of the slope field
$u$ and the height field $h$ (arbitrary units).
}
\end{figure}
%%%%%%%%%%%%%%%%%%%%%
%%%%%%%%%%%%%%%%%%%%%
%%%%%%%%%%%%%%%%%%%%%
\begin{figure}
\centerline
{
\epsfxsize=14cm
\epsfbox{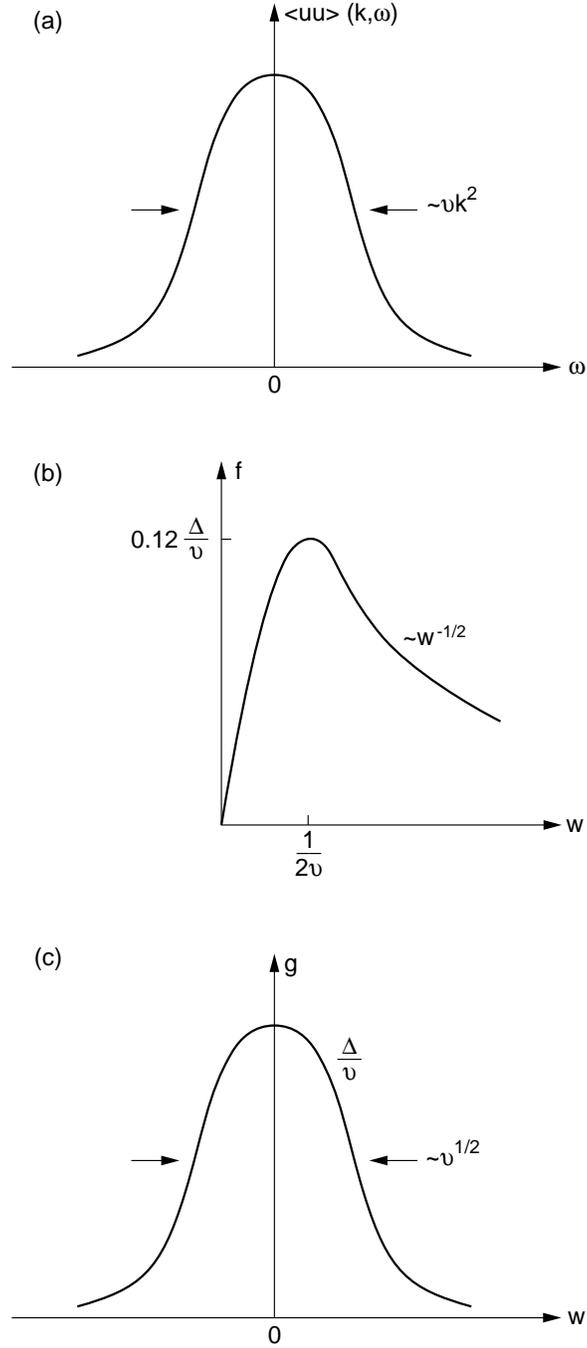}
}
\caption{
In a) we depict  the slope correlation function for the
diffusive mode in the linear EW case. The Lorentzian is centered about
$\omega = 0$ and has the ``hydrodynamical'' line width  $\nu k^2$
vanishing in the long wave length limit. In b) we show the scaling
functions $f$ for the space and time-dependent slope correlations.
For large $w$ $f$ falls off as $w^{-1/2}$, for small $w$ 
$f\rightarrow 0$ with an essential singularity for $w=0$. $f$
is peaked at the value $\sim\text{0.12}\Delta/\nu$ for
$w=1/2\nu$. In b) we have shown the scaling function $g$ for
the wave number-frequency dependent correlations. $f$ has a
Lorentzian form with height $\Delta/\nu$ and width
$\sim\nu^{1/2}$ (arbitrary units).
}
\end{figure}
%%%%%%%%%%%%%%%%%%%%%
%%%%%%%%%%%%%%%%%%%%%
%%%%%%%%%%%%%%%%%%%%%
\begin{figure}
\centerline
{
\epsfxsize=15cm
\epsfbox{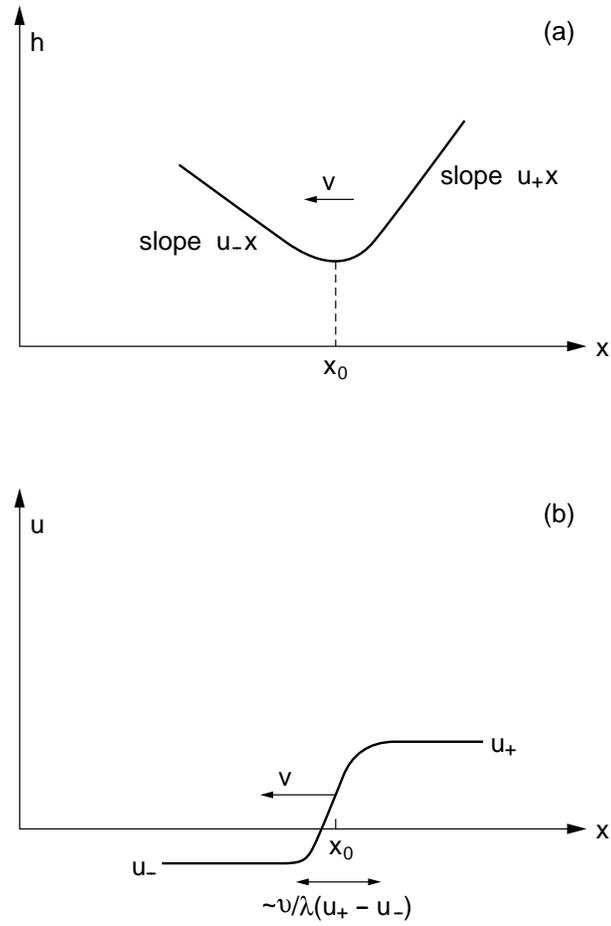}
}
\caption{
We show a single moving soliton profile propagating to the left
and the corresponding smoothed cusp in the growth profile.
This configuration is driven by currents at the boundaries,
corresponding to non-vanishing $u_\pm$ and is persistent in time
(arbitrary units).
}
\end{figure}
%%%%%%%%%%%%%%%%%%%%%
%%%%%%%%%%%%%%%%%%%%%
%%%%%%%%%%%%%%%%%%%%%
\begin{figure}
\centerline
{
\epsfxsize=15cm
\epsfbox{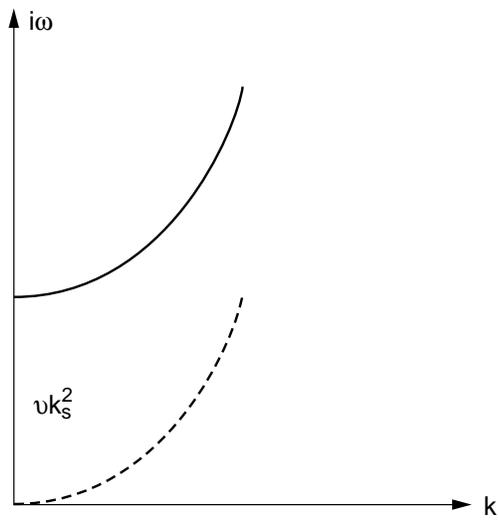}
}
\caption{
We show the diffusive dispersion law in the presence of
a soliton. The gap in the spectrum is given by
$\nu k_s^2 = \lambda^2 u_+^2/4\nu$ where $u_+$ is the soliton
amplitude. The dashed line indicates the gapless spectrum in
the linear case (arbitrary units).
}
\end{figure}
%%%%%%%%%%%%%%%%%%%%%
%%%%%%%%%%%%%%%%%%%%%
%%%%%%%%%%%%%%%%%%%%%
\begin{figure}
\centerline
{
\epsfxsize=15cm
\epsfbox{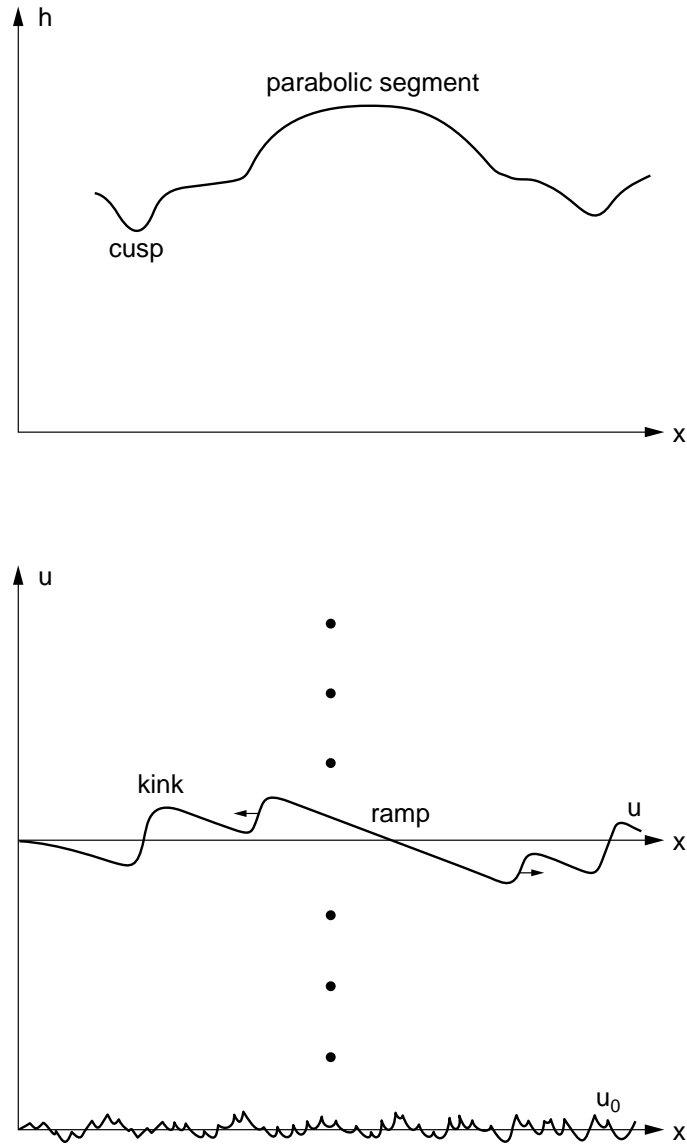}
}
\caption{
We here depict the transient evolution of the slope field
$u$ from an initial configuration $u_0$ in the case of the noiseless
Burgers equation. We have also shown the evolution of the associated
height field $h$. The transient morphology
consists of propagating {\em right hand} solitons connected by
ramps (arbitrary units).
}
\end{figure}
%%%%%%%%%%%%%%%%%%%%%
%%%%%%%%%%%%%%%%%%%%%
%%%%%%%%%%%%%%%%%%%%%
\begin{figure}
\centerline
{
\epsfxsize=15cm
\epsfbox{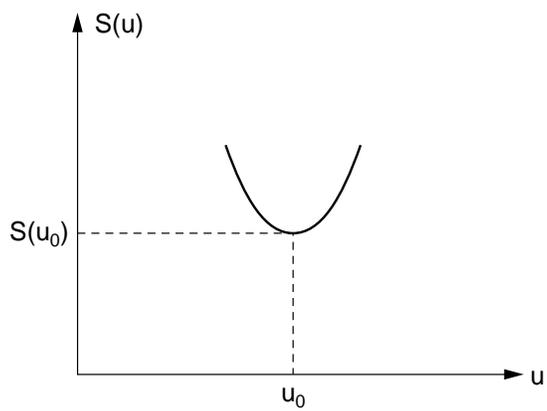}
}
\caption{
Here we depict in graphic form the basic principle of
an asymptotic steepest descent or saddle point calculation. The leading
contribution to the integral $I(\Delta)$ in Eq. (\ref{int}) arises from
the saddle point $u_0$ (in the figure a minimum) and nearby fluctuations
$\delta u =u_0 - u$ (arbitrary units).
}
\end{figure}
%%%%%%%%%%%%%%%%%%%%%
%%%%%%%%%%%%%%%%%%%%%
%%%%%%%%%%%%%%%%%%%%%
\begin{figure}
\centerline
{
\epsfxsize=15cm
\epsfbox{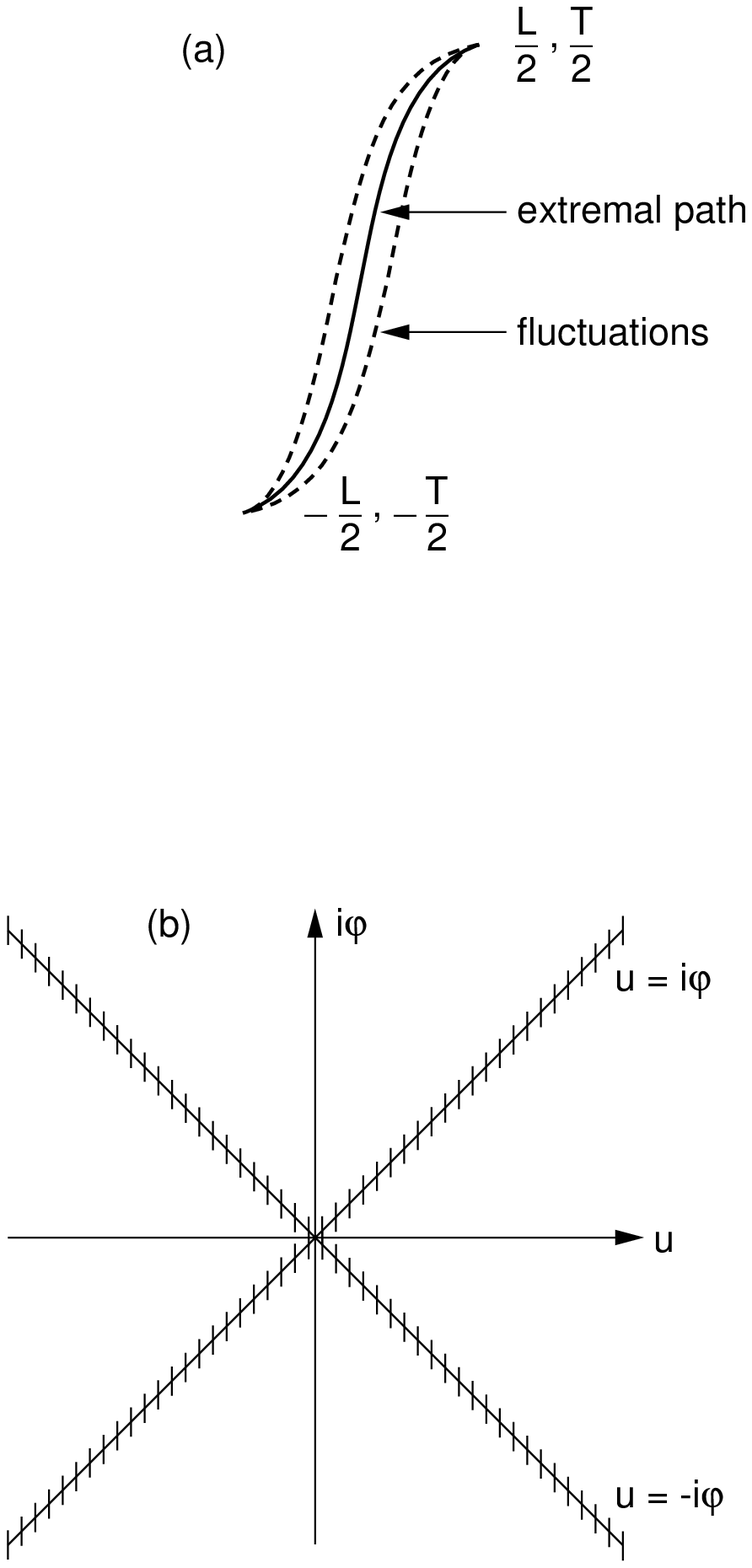}
}
\caption{
In a) we show the ``classical'' path corresponding to
the stationary point of the action $S$ in
the weak noise limit and nearby paths corresponding to fluctuations.
In b) we show the saddle point regions in
$(u,\varphi)$ phase space (arbitrary units).
}
\end{figure}
%%%%%%%%%%%%%%%%%%%%%
%%%%%%%%%%%%%%%%%%%%%
%%%%%%%%%%%%%%%%%%%%%
\begin{figure}
\centerline
{
\epsfxsize=15cm
\epsfbox{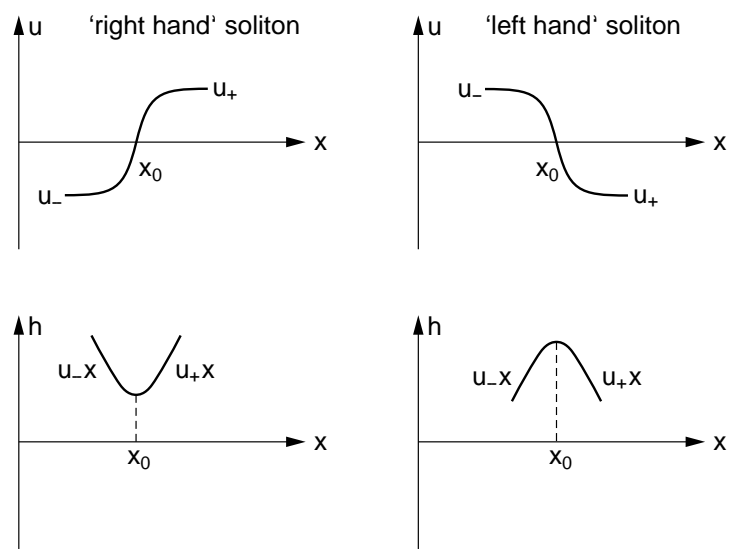}
}
\caption{
In a) and b) we show the static {\em right} and {\em left hand} solitons
and the smoothed static downward and upward cusps
in the associated height field
(arbitrary units).
}
\end{figure}
%%%%%%%%%%%%%%%%%%%%%
%%%%%%%%%%%%%%%%%%%%%
%%%%%%%%%%%%%%%%%%%%%
\begin{figure}
\centerline
{
\epsfxsize=15cm
\epsfbox{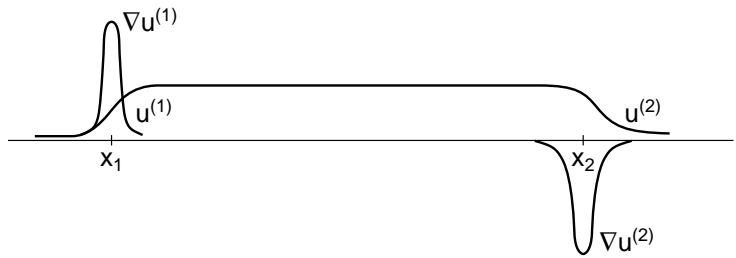}
}
\caption{
We show the overlap between two well-separated solitons
(arbitrary units).
}
\end{figure}
%%%%%%%%%%%%%%%%%%%%%
%%%%%%%%%%%%%%%%%%%%%
%%%%%%%%%%%%%%%%%%%%%
\begin{figure}
\centerline
{
\epsfxsize=15cm
\epsfbox{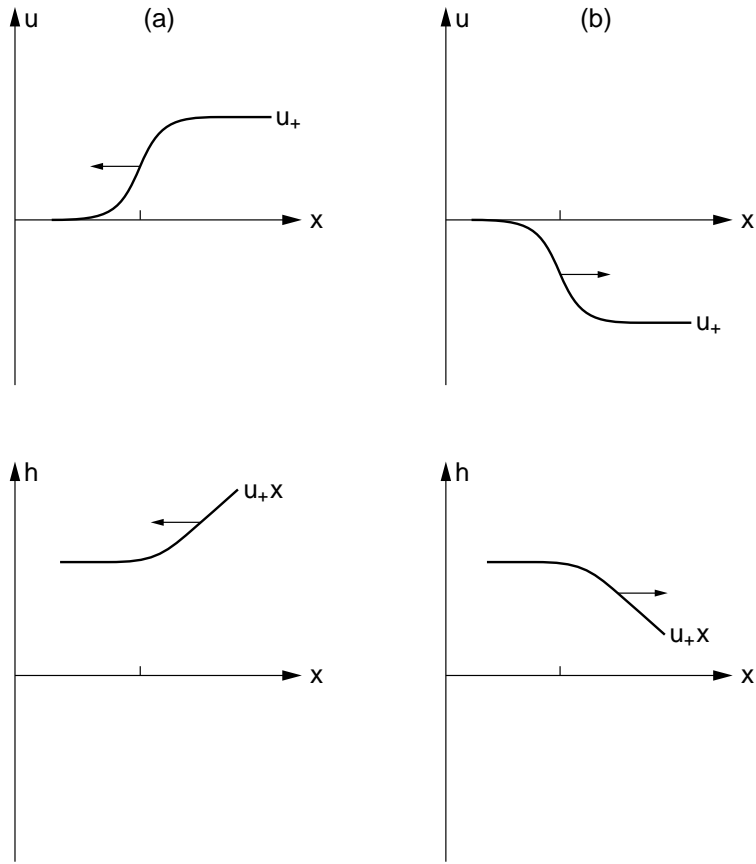}
}
\caption{
In a) and b) we show  right and left-moving solitons with vanishing
amplitude at $x=-L/2$. The associated height profiles correspond
to the bottom and top part of a step, respectively
(arbitrary units)..
}
\end{figure}
%%%%%%%%%%%%%%%%%%%%%
%%%%%%%%%%%%%%%%%%%%%
%%%%%%%%%%%%%%%%%%%%%
\begin{figure}
\centerline
{
\epsfxsize=15cm
\epsfbox{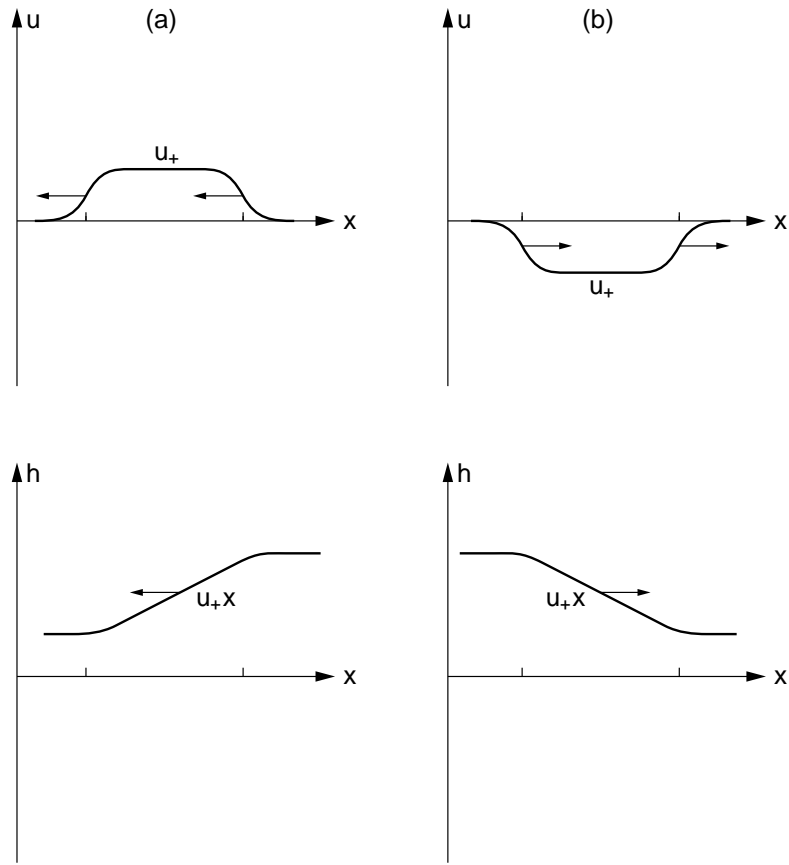}
}
\caption{
In a) and b) we show two two-soliton configurations moving
with opposite
velocities corresponding to the left and right propagation of
a step in the height profile (arbitrary units).
}
\end{figure}
%%%%%%%%%%%%%%%%%%%%%
%%%%%%%%%%%%%%%%%%%%%
%%%%%%%%%%%%%%%%%%%%%
\begin{figure}
\centerline
{
\epsfxsize=15cm
\epsfbox{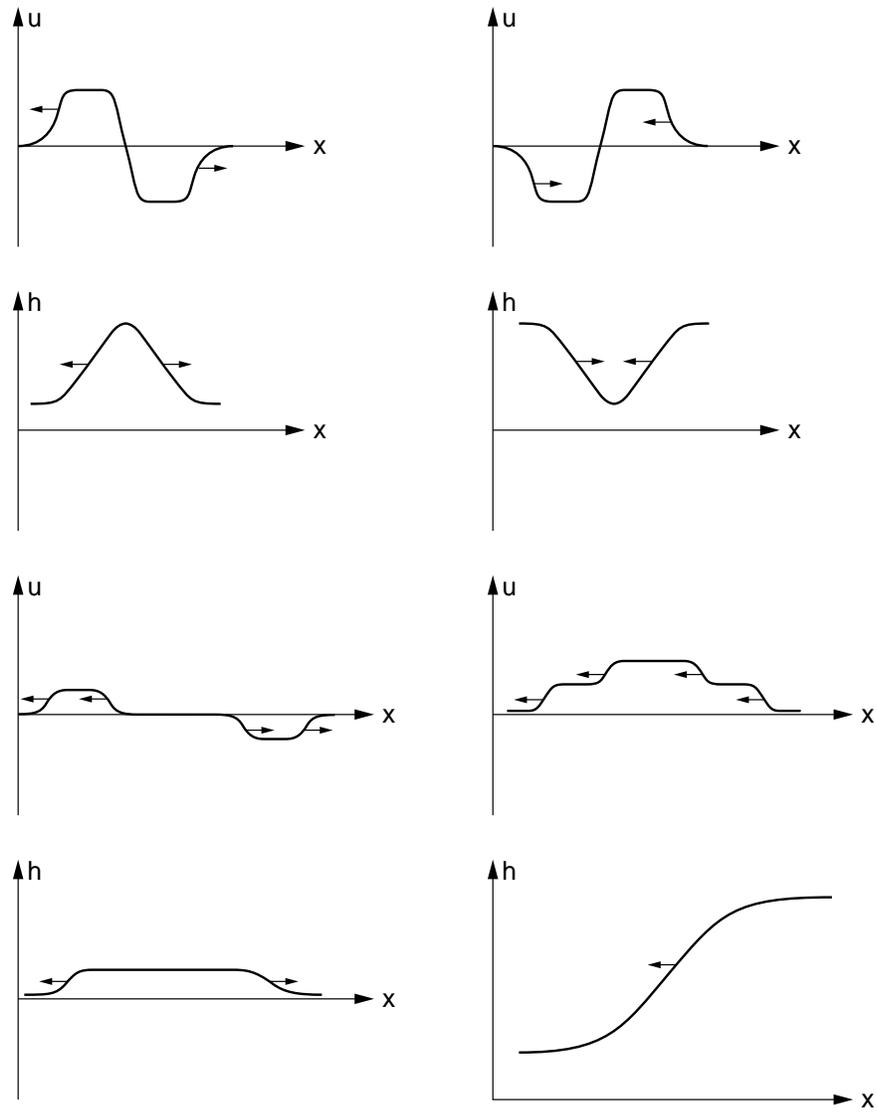}
}
\caption{
We show the soliton configuration corresponding to a)
the growth of a tip, b) the filling in of an indentation,
c) the growth of a plateau, and d) the ``renormalization'' of a step
(arbitrary units).
}
\end{figure}
%%%%%%%%%%%%%%%%%%%%%
%%%%%%%%%%%%%%%%%%%%%
%%%%%%%%%%%%%%%%%%%%%
\begin{figure}
\centerline
{
\epsfxsize=15cm
\epsfbox{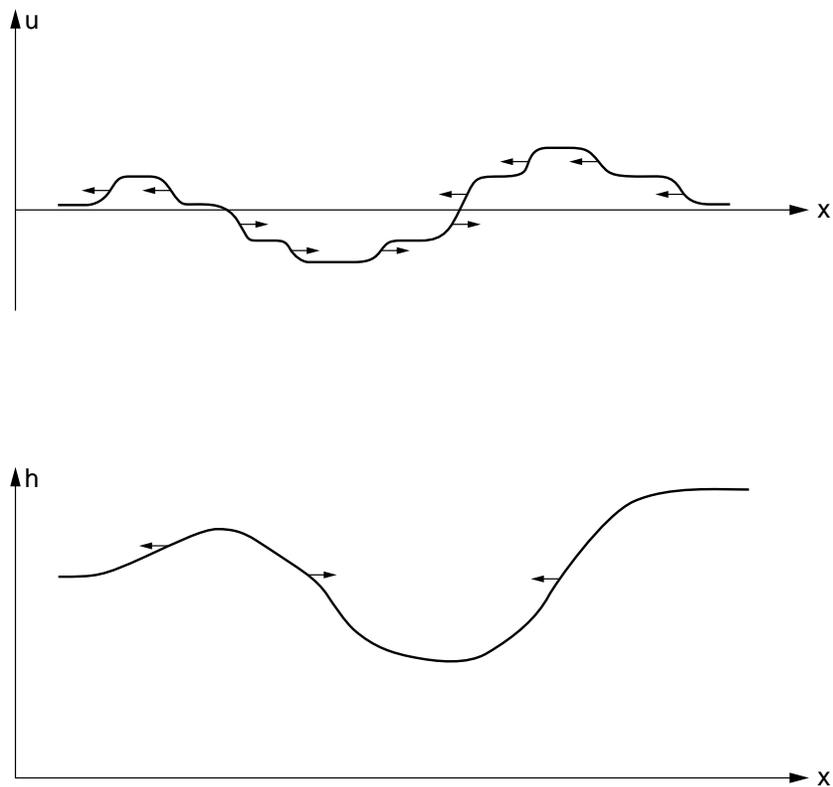}
}
\caption{
The general growth of an interface in terms of a dilute gas of
solitons (arbitrary units).
}
\end{figure}
%%%%%%%%%%%%%%%%%%%%%
%%%%%%%%%%%%%%%%%%%%%
%%%%%%%%%%%%%%%%%%%%%
\begin{figure}
\centerline
{
\epsfxsize=15cm
\epsfbox{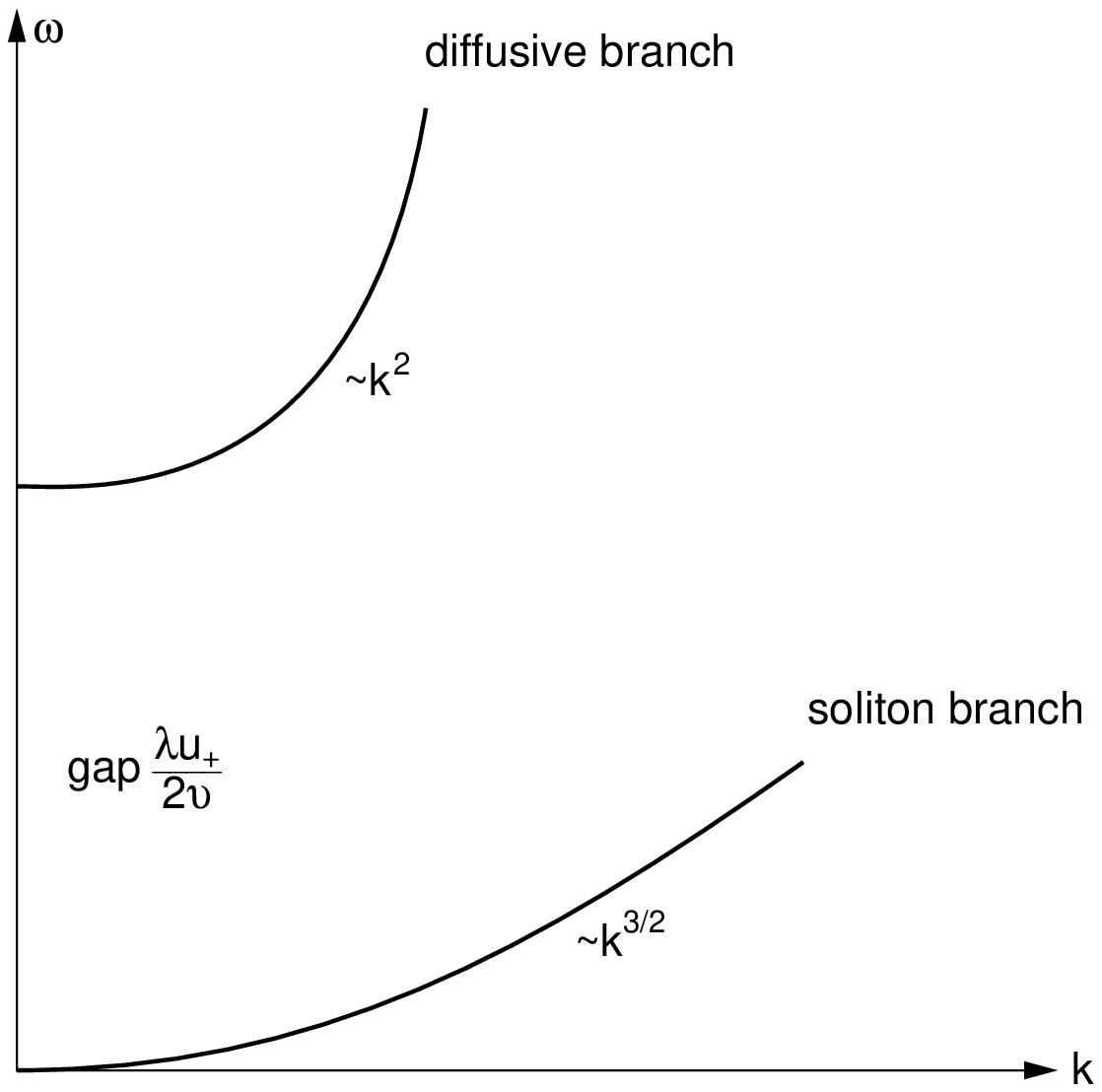}
}
\caption{
We depict the quadratic diffusive dispersion law with  gap
$\lambda^2 u_+^2/4\nu$ and the gapless
soliton dispersion law with fractional exponent $3/2$
(arbitrary units).
}
\end{figure}
%%%%%%%%%%%%%%%%%%%%%

\begin{references}
\bibitem{footnote2}
More precisely, the determinant can be shown to have the
form \cite{zinn-justin}
$J = \exp{[-(1/2)Tr_{xt}[\nu\nabla^2+\lambda( u\nabla u +\nabla u)]]}$
and thus contributes an additional term $G_J$ to the action $G$.
However, as discussed in \cite{dedom1,dedom2,dedom3,janssen2}
the term $G_J = \log{J}$ precisely cancels acausal contributions
in a perturbation expansion in $\lambda$, i.e., closed loops.
This also follows from a direct expansion of the Langevin
equation (\ref{bur2}) in powers of $\lambda$ combined with succesive
noise contractions. Here loops are absent due to the causal character
of the unperturbed propagator $(\partial/\partial t - \nu\nabla^2)^{-1}$.
Thus with this proviso we are entitled to use the identity (\ref{id}).
\bibitem{footnote3}
The connection between the Fokker Planck description of the
noisy Burgers equation \cite{halpin,fogedby3}, entailing the
stationary distribution $P(u)$, and the present path integral
formulation is clearly of interest and will be pursued in
another context; otherwise, we refer to \cite{zinn-justin}
for a general discussion.
\bibitem{footnote4}
The choice of Poisson algebra is to some extent arbitrary. 
The present choice has been dictated by the analogy between the present
``stochastic'' path integral and the usual Feynman phase space integral.
\bibitem{footnote5} 
In writing down the ``quantum Hamiltonian'' $\hat{H}$ we have 
chosen a normal ordering of the operators $u$ and $\varphi$ in
the interaction term, i.e., we have placed the ``momentum'' $u$ to
the left of the ``coordinate'' $\varphi$. In the present heuristic 
discussion of the ``quantum mechanics'' the ordering is,
however, immaterial; otherwise see \cite{zinn-justin,das}.
\bibitem{footnote6}
In writing down the Heisenberg equations of motion we have in the
usual way gone from the Schr\"{o}dinger picture with tine evolution
operator $\hat{U}(t)=\exp{(-i(\Delta/\nu)\hat{H}t)}$ to the Heisenberg
picture with operator time evolution 
$\hat{u}(t)=
\exp{(+i(\Delta/\nu)\hat{H}t)}\hat{u}(0)\exp{(-i(\Delta/\nu)\hat{H}t)}$;
note that this transformation does not require $\hat{H}$ to be Hermitian
or $\hat{U}(t)$ to be unitary.
\bibitem{footnote7}
We note that the Heisenberg equations of motion ``drive the operators 
away from hermiticity'' since $\hat{H}$ is non-Hermitian.
\bibitem{footnote8}
Strictly speaking, since the Hamiltonian is non-Hermitian,
we must define ``right hand'' and ``left hand'' 
eigenstates of $\hat{H}$.
\end{references}
\end{document}